\documentclass[fleqn,usenatbib,letterpaper]{mnras}
\usepackage[T1]{fontenc}
\usepackage{ae,aecompl}
\usepackage{amsmath}
\usepackage{amssymb}
\usepackage{graphicx}
\graphicspath{{figure/}}

\usepackage{url}
\newcommand{\kms}{\,km\,s$^{-1}$}
\newcommand{\Msun}{\,M$_\odot$}
\newcommand{\sqarcmin}{\,arcmin$^{2}$}
\newcommand{\sqdeg}{\,deg$^{2}$}

\newcommand{\molh}{$\text{H}_2$}
\newcommand{\twosqdeg}{2-deg$^{2}$}
\newcommand{\cubeMpc}{\,Mpc$^{3}$}

\defcitealias{Bruzual2003}{BC03}
\defcitealias{Gnedin2011}{GK}
\defcitealias{Bigiel2008}{Big}

\title[SAM forecasts for Roman deep-wide surveys]{Semi-analytic forecasts for \textit{Roman} -- the beginning of a new era of deep-wide galaxy surveys}

\author[L. Y. A. Yung et al.]{L. Y. Aaron\ Yung,$^{1}$\thanks{E-mail: aaron.yung@nasa.gov}\thanks{NPP Fellow}
Rachel S.\ Somerville,$^{2}$
Steven L.\ Finkelstein,$^{3}$
\newauthor
Peter\ Behroozi,$^{4,5}$
Romeel\ Dav\'e,$^{6, 7, 8}$
Henry C.\ Ferguson,$^{9}$
Jonathan P.\ Gardner,$^{1}$
\newauthor
Gerg\"o\ Popping,$^{10}$
Sangeeta\ Malhotra,$^{1}$
Casey\ Papovich,$^{11,12}$
James E.\ Rhoads,$^{1}$
\newauthor
Micaela B.\ Bagley,$^{3}$
Michaela\ Hirschmann,$^{13}$
and Anton M. Koekemoer$^{9}$
\\
$^{1}$Astrophysics Science Division, NASA Goddard Space Flight Center, 8800 Greenbelt Rd, Greenbelt, MD 20771, USA\\
$^{2}$Center for Computational Astrophysics, Flatiron Institute, 162 5th Ave, New York, NY 10010, USA\\
$^{3}$Department of Astronomy, The University of Texas at Austin, Austin, TX 78712, USA\\
$^{4}$Department of Astronomy, University of Arizona, 933 N Cherry Ave, Tucson, AZ 85721, USA\\
$^{5}$Division of Science, National Astronomical Observatory of Japan, 2-21-1 Osawa, Mitaka, Tokyo 181-8588, Japan\\
$^{6}$Institute for Astronomy, University of Edinburgh, Edinburgh EH9 3HJ, UK\\
$^{7}$Department of Physics and Astronomy, University of the Western Cape, Cape Town 7535, South Africa\\
$^{8}$South African Astronomical Observatory, Cape Town 7925, South Africa\\
$^{9}$Space Telescope Science Institute, 3700 San Martin Drive, Baltimore, MD 21218, USA\\
$^{10}$European Southern Observatory, Karl-Schwarzschild-Strasse 2, D-85748 Garching, Germany\\
$^{11}$Department of Physics and Astronomy, Texas A\&M University, College Station, TX 77843, USA\\
$^{12}$George P. and Cynthia Woods Mitchell Institute for Fundamental Physics and Astronomy, Texas A\&M University, \\\;\;\;\,\,College Station, TX 77843, USA\\
$^{13}$Institute of Physics, Laboratory of Galaxy Evolution, Ecole Polytechnique F\'ede\'erale de Lausanne (EPFL), \\\;\;\;\;Observatoire de Sauverny, 1290 Versoix, Switzerland
}
\date{Accepted XXX. Received YYY; in original form ZZZ}

\pubyear{2022}

\begin{document}
\label{firstpage}
\pagerange{\pageref{firstpage}--\pageref{lastpage}}
\maketitle

\begin{abstract}
The \emph{Nancy Grace Roman Space Telescope}, NASA's next flagship observatory, will redefine deep-field galaxy survey with a field of view two orders of magnitude larger than \emph{Hubble} and an angular resolution of matching quality.
These future \textit{deep-wide} galaxy surveys necessitate new simulations to \textit{forecast} their scientific output and to optimise survey strategies.
In this work, we present five realizations of \twosqdeg\ lightcones, containing a total of $\gtrsim25$ million simulated galaxies with  $-16\gtrsim M_\text{UV}\gtrsim-25$ spanning $z\sim0$ to 10.
This dataset enables a new set of experiments with the impacts of survey size on the derived galaxy formation and cosmological constraints.
The intrinsic and observable galaxy properties are predicted using a well-established, physics-based semi-analytic modelling approach. We provide forecasts for number density, cosmic SFR, field-to-field variance, and angular two-point correlation functions, and demonstrate how the future wide-field surveys will be able to improve these measurements relative to current generation surveys.
We also present a comparison between these lightcones and others that have been constructed with empirical models.
The mock lightcones are designed to facilitate the exploration of multi-instrument synergies and connecting with current generation instruments and legacy surveys.
In addition to \emph{Roman}, we also provide photometry for a number of other instruments on upcoming facilities, including Euclid and Rubin, as well as the instruments that are part of many legacy surveys.
Full object catalogues and data tables for the results presented in this work are made available through a web-based, interactive portal.
\end{abstract}
%% 250 words

%%%%%%%%%%%%%%%%%%%%%%%%%%%%%%%%%%%%%%%%%%%%%%%%%%
%%%%%%%%%%%%%%%%% BODY OF PAPER %%%%%%%%%%%%%%%%%%

\section{Introduction}

Over the past three decades, observations with the \emph{Hubble Space Telescope} have revolutionized our understanding of the assembly histories of galaxies in the context of the Universe's overall evolutionary history.
The multi-cycle treasury program Cosmic Assembly Near-infrared Deep Extragalactic Legacy Survey \citep[CANDELS;][]{Grogin2011, Koekemoer2011} has established a handful of relatively well-surveyed legacy fields driven mainly by \emph{Hubble} in conjunction with the \emph{Spitzer Space Telescope} and many ground-based telescopes.
These surveys reliably reach a 5$\sigma$ depth of $\sim26.5$ (with some variation across different fields).
The five CANDELS legacy fields combined cover a total of $\sim850$\,\sqarcmin, reaching galaxies as far as $z\sim11$ \citep[e.g.][]{Tacchella2021, Finkelstein2021}.
Within the coverage of the CANDELS fields, the \emph{Hubble} Ultra Deep Field \citep[HUDF;][]{Beckwith2006, Ellis2013, Koekemoer2013} and the eXtreme Deep Field \citep[XDF;][]{Illingworth2013, Oesch2013, Oesch2018} have pioneered imaging the extremely deep universe, reaching a 5$\sigma$ depth of $\sim29.8$, at the expense of long exposure times.

The \emph{Nancy Grace Roman Space Telescope}\footnote{\url{https://roman.gsfc.nasa.gov}}, or \emph{Roman} in short, NASA's next premier space-based observatory, is expected to survey the Universe at unprecedented efficiency with its extreme wide-field survey capability \citep{Spergel2013, Spergel2015}.
As of today, the \emph{Roman} mission has been confirmed by NASA in March 2020, passed various critical design reviews in 2021, and is on track towards an anticipated launch date before May 2027.
\emph{Roman}'s advanced optical system and its onboard Wide Field Instrument (WFI) together offer a field of view of $\sim0.28$\sqarcmin, which is approximately a hundred times bigger than that of \emph{Hubble}, and possesses an infrared sensitivity that is comparable to or exceeding that of \emph{Hubble} \citep{Pasquale2014, Pasquale2018}.
In other words, the size of a single \emph{Roman} WFI pointing is comparable to the total area of the five legacy CANDELS fields combined.
In this new era of deep-field galaxy surveys, \textit{every Roman field is a wide field when compared to current generation observations.}

\emph{Roman}'s wide-field capabilities will enable coverage over larger survey areas with many fewer pointings, which effectively permits longer exposure over the targeted fields and increases the survey depths relative to past \emph{Hubble} galaxy surveys with similar survey size and allocated time.
Therefore, the \textit{new generation} of deep-field galaxy surveys enabled by \emph{Roman} is expected to reach (or even surpass) the depth of HUDF over areas that are orders of magnitude larger than current generation CANDELS fields (see the \emph{Roman} Ultra Deep Field concept described in
\citealt{Koekemoer2019}).
Furthermore, \emph{Roman} will, for the first time, enable high-redshift (e.g. z > 6) surveys spanning multiple square degrees that reach depths comparable to \emph{Hubble} extragalactic legacy surveys (e.g. CANDELS).
Current surveys at this scale are driven largely by ground-based instruments, such as the Spitzer/HETDEX Exploratory Large Area survey \citep[SHELA;][]{Papovich2016, Stevans2018, Stevans2021, Wold2019}, the Great Optically Luminous Dropout Research Using Subaru HSC survey \citep[GOLDRUSH;][]{Harikane2018, Harikane2021, Ono2018, Toshikawa2018}, and the Lyman Alpha Galaxies in the Epoch of Reionization survey \citep[LAGER;][]{Zheng2017, Hu2019a, Wold2022}.
These large surveys by ground-based telescopes have identified millions of sources up to $z\sim7$, providing robust constraints for the number density and clustering statistics of bright, massive objects.
However, ground-based instruments are subject to various disadvantages compared to their space-based counterparts and are less ideal for high-redshift explorations.
Future \emph{Roman} surveys will provide higher angular resolution images, and reach depths that are not accessible from the ground, particularly in the near-IR.
These surveys are expected to deliver robust statistical constraints on the physical properties and spatial distributions of galaxies across cosmic time and on the clustering of galaxies, which have strong implications for the nature of dark matter and the formation of large-scale structure.
Furthermore, these large survey areas will be crucial for robustly constraining the number density of rare, luminous galaxies at extreme redshifts \citep{Harikane2021a}.

While \emph{JWST} \citep{Gardner2006}, NASA's latest space-based observatory, remains the most sensitive infrared telescope in operation and will be the powerhouse for ultra-deep surveys (e.g. $m_\text{AB} > 29$), its relatively narrow field of view is not expected to significantly increase survey area compared to past \emph{Hubble} surveys.
Wide-field \emph{JWST} programs, such as Cycle 1 GO/Treasury program COSMOS-Web \citep[][$\sim0.60$\,\sqdeg]{Kartaltepe2021, Casey2022}, will still be able to cover areas that are comparable to a handful of WFI pointings and are therefore an important pathfinder to inform future Roman deep-field survey strategies.
Thanks to the excellent launch delivered by ESA and Arianespace that resulted in significant conservation of \emph{JWST}'s on-board fuel, its expected mission lifespan has been prolonged to upwards of 15 to 20 years. This implies that \emph{JWST} is expected to overlap significantly with NASA's \emph{Roman}, NSF's \emph{Vera C. Rubin Observatory}, and ESA's \emph{Euclid} mission. Therefore, the synergy across \emph{JWST} and these flagship observatories plays a crucial part in increasing the scientific productivity of all facilities. For instance, \textit{JWST} wide-field surveys are important pathfinders for future \textit{deep-wide} surveys that are expected to be conducted by these next generation wide-field survey instruments. \emph{JWST}'s superb sensitivity, mid-IR coverage, and high-resolution spectroscopic capabilities are also suitable for follow-up investigations complementary to the wide-field surveys.

In anticipation of this new generation of deep-wide surveys, predictions or \textit{forecasts} of galaxy population properties over large volumes are essential for the assessment and development of optimal survey strategies, studying synergies between planned observations with different facilities, and ultimately realizing the full scientific potential of these observations. There are three main approaches for modelling galaxy formation: numerical hydrodynamic simulations, semi-analytic models, and empirical models. The first two approaches are similar in that they are based on a set of \emph{a priori} physical processes, such as gas cooling and inflow, star formation and stellar feedback, chemical enrichment, and black hole growth and feedback \citep{Somerville2015a, Naab2017}. In empirical approaches, either a mapping is created between dark matter halos and galaxy properties using observational constraints, as in sub-halo abundance matching models \citep[SHAMs,][]{Wechsler2018}, or models are constructed based purely on existing observations \citep[e.g. JADES,][]{Williams2018}. Numerical hydrodynamic simulations provide very detailed predictions, but it is currently very challenging or impossible to run simulations with volumes comparable to the anticipated next generation of wide-field surveys. Empirical models can very efficiently populate large volumes, but they have limited predictive or interpretive power, and are not able to produce self-consistent predictions for multiple galaxy components. Semi-analytic models are based on physical processes set within a cosmological framework, and self-consistently predict multiple components of galaxies (such as stars, metals, dust, and different phases of gas) \citep{Somerville1999, Somerville2015, Benson2010, Cowley2018, Henriques2015a, Henriques2020}.
At the same time, they are computationally efficient enough to be able to explore parameter space, and create forecasts for relatively large volumes.
In the past, this type of theoretical framework has been shown to be a valuable tool that adds scientific value to high-redshift galaxy surveys, including the interpretation of the CANDELS surveys \citep{Somerville2021} and can facilitate the planning of future \emph{JWST} survey programs, such as CEERS\footnote{The \textit{Cosmic Evolution Early Release Science} Survey \citep{Finkelstein2017, Finkelstein2022a, Finkelstein2022b, Bagley2022a}}, NGDEEP\footnote{The \textit{Next Generation Deep Extragalactic Exploratory Public} Survey \citep{Finkelstein2021a}}, and PRIMER\footnote{The \textit{Public Release IMaging for Extragalactic Research} Survey \citep{Dunlop2021}} \citep{Yung2022}.
These models are also useful for cross-correlating with a large set of anticipated results from intensity mapping surveys, such as EXCLAIM \citep[e.g.][]{Switzer2021, Yang2021a, Pullen2022}.
Useful predictions for \textit{JWST} have also been made by other groups using semi-analytic models \citep[e.g.][]{Dayal2014, Dayal2015, Cowley2018} and hydrodynamic cosmological simulations \citep[e.g.][]{Vogelsberger2020a, Kannan2022, Kannan2021, Wilkins2022, Wilkins2022a}.

Lightcones are an effective tool for bridging the gap between simulations and observations. Conventional numerical techniques are carried out by tracking the positions and velocities of mass particles within a simulated cubical (co-moving) volume and provide predictions of galaxy properties in snapshots at discrete output times. However, of course when we observe the Universe, we observe galaxies along a past lightcone.
Based on snapshots output by numerical simulations, galaxies and halos within a cubical simulated volume can be sampled along such a past lightcone, and mock catalogs constructed in this way are commonly referred to as \textit{lightcones}.
Lightcones can be constructed based on hydrodynamic simulations \citep[e.g.][]{Snyder2017, Snyder2022} or with halos extracted from $N$-body simulations and filled in with galaxies from empirical models \citep[e.g.][]{Behroozi2019, Drakos2021} or semi-analytic models \citep{Overzier2013, Bernyk2016, Smith2017, Barrera2022}.
We note that this is a broad overview of lightcone construction and refer the reader to the above referenced works for detailed descriptions.

While lightcones extracted from hydrodynamic simulations contain spatially resolved galaxies that are tracked self-consistently within the simulated environment, the size and mass resolution of these lightcones are limited due to the relatively high computational expense of the underlying hydro simulations. On the other hand, semi-analytic models and empirical methods built on top of cheaper dark matter only simulations provide a more cost effective alternative, enabling larger area mock fields to be simulated.
In anticipation of upcoming multi-\sqdeg\ \textit{deep-wide} surveys, we present physically-based mock catalogues with tens of millions of galaxies constructed based on the halos from the Bolshoi-Planck simulation \citep{Klypin2016} and galaxies predicted by the physically-motivated Santa Cruz SAM \citep{Somerville2015}.

This work is built based on the well-established Santa Cruz Semi-analytic models \citep{Somerville1999, Somerville2015}, which have been extensively compared with observations \citep{Somerville2015, Yung2019, Yung2019a, Yung2021} and with the predictions of numerical hydrodynamic simulations \citep{Pandya2020, Gabrielpillai2021}. In particular, this work builds on the techniques presented in \citet{Somerville2021} and the \textit{Semi-analytic forecasts for JWST} paper series \citep{Yung2019,Yung2019a,Yung2020,Yung2020a,Yung2021}. In these works, the SC SAM modelling framework was shown to reproduce existing constraints on physical and observable properties of high-redshift galaxies ($4<z<10$) and AGN ($2<z<7$), as well as their subsequent impact on the cosmic hydrogen and helium reionization history.
The final paper of the series \citep{Yung2022} presented a large collection of wide-field ($\sim1000$\;\sqarcmin) and ultra-deep (rest-frame $M_\text{UV} \lesssim -12$) lightcones and associated data products.
In this complementary paper, we present a set of \twosqdeg\ lightcones, with the aim of providing physically accurate predictions for the large-scale distribution and clustering of galaxies. Using these predictions, we present quantitative predictions for the expected uncertainty due to field-to-field variance in both one-point distributions (object counts) and two-point statistics (two-point correlation functions).
In addition to \emph{Roman}, \emph{Euclid}, and \emph{Rubin}, the dataset also includes photometric bands presented in past \emph{CANDELS} and \emph{JWST} mock catalogues, and can be used to explore the synergy across \emph{Roman}, \emph{JWST}, \emph{Hubble}, \emph{Spitzer}, and many ground-based observatories.
In this work series, we use two NASA flagships, \emph{Roman} and \emph{JWST} as practical examples to demonstrate how these predictions can be used. The physically motivated predictions made with the Santa Cruz semi-analytic model can be easily adapted to make predictions for other space- and ground-based facilities.

All mock catalogues and simulated results presented in this work series are accessible through the project homepage\footnote{\url{https://www.simonsfoundation.org/semi-analytic-forecasts/}} and the Flatiron Institute Data Exploration and Comparison Hub (Flathub\footnote{\url{https://flathub.flatironinstitute.org/group/sam-forecasts}}).

The key components of this work are summarized as follows:
we provide a concise summary of the galaxy formation model and present the simulated lightcones in Sections \ref{sec:models} and \ref{sec:data_release}, respectively.
We present the main results in Section \ref{sec:results}.
We discuss our findings in Section \ref{sec:discussion}, and a summary and conclusions follow in Section \ref{sec:snc}.

\section{Lightcone Construction Pipeline with Physical Models}
\label{sec:models}

In this section, we provide a concise summary of the semi-analytic model (SAM) for galaxy formation developed over the years by the Santa Cruz group and collaborators \citep{Somerville1999, Somerville2008, Somerville2012, Somerville2015, Popping2014}. We refer the reader to these papers for a full description of the model components. The specific models and configurations for galaxies and AGN are documented in \citet{Yung2019,Yung2021} and \citet{Somerville2021}. Free parameters in these models are calibrated as described in \citet{Yung2019} and \citet{Somerville2021}. Throughout this work, we adopt cosmological parameters $\Omega_\text{m} = 0.308$, $\Omega_\Lambda = 0.692$, $H_0 = 67.8$ \kms Mpc$^{-1}$, $\sigma_8$ = 0.831, and $n_s = 0.9665$; which are broadly consistent with the ones reported by the Planck Collaboration in 2015 (Planck Collaboration XIII \citeyear{Planck2016}) and are consistent with the rest of the paper series.
All magnitudes presented in this work are expressed in the AB system \citep{Oke1983} and all uses of log are base 10 unless otherwise specified.
The calculations in this work are carried out with \texttt{ASTROPY} \citep{Robitaille2013,Price-Whelan2018}, \texttt{NUMPY} \citep{vanderWalt2011}, \texttt{SCIPY} \citep{Virtanen2020}, and \texttt{pandas} \citep{Reback2022}.
We provide the code that we used to calculate the co-moving volume of the lightcone slice in Appendix \ref{appendix:b}. This simple calculation is essential to deriving many volume-averaged quantities presented in this work.

\subsection{Dark matter cones and merger histories constructions}
\label{sec:darkcon}
The set of five realizations of \twosqdeg\ lightcones presented in this work are constructed with the same process detailed in \citet{Somerville2021} and \citet{Yung2022}. The dark matter halos that form the basis of the \twosqdeg\ lightcones are sourced from the Small MultiDark-Planck (SMDPL) simulation from the MultiDark simulation suite \citep{Klypin2016}.
This dark matter-only $N$-body simulation has a volume of (400 Mpc $h^{-1}$)$^3$ and dark matter particle mass of $M_\text{DM} \sim 9.6\times10^7$ \Msun\,$h^{-1}$.
Here $h$ denotes $h_{100} \equiv H_{0}/100$.
Halos in this cosmological simulation are identified using the six-dimensional phase-space halo finder \textsc{rockstar} and \textsc{consistent trees} \citep{Behroozi2013b, Behroozi2013c}.
We refer the reader to \citet{Rodriguez-Puebla2016} for details on the halo catalogues.
As in that work, we adopt the halo virial mass definition from \citet{Bryan1998}.
The mass threshold for resolved halos is set to $M_\text{res} \sim 10^{10}$ \Msun, which is equivalent to the mass of $\sim 100$ dark matter particles.

The dark matter halos in SMDPL are then arranged into mock observed fields spanning \twosqdeg\ each between $0 \lesssim z \lesssim 10$, using the \texttt{lightcone} package that is released as part of \textsc{UniverseMachine} \citep{Behroozi2019}.
For each lightcone realization, the lightcone tool picks a random origin and viewing angle within the base dark matter-only simulation (see Table 1), and includes all halos that fall within the specified survey area. The tool makes use of the periodic boundary conditions when halos lie beyond the boundary of the simulated volume. The distance along the lightcone axis determines the redshift of the simulation snapshot from which halo properties are taken. While the lightcones in this paper were allowed to pass through the same region of the simulation volume multiple times, since the halos are sampled at a random angle, it is unlikely that a slice of the lightcone will be repeated in the same redshift slice (which happens only if halos are sampled in a slice that is perpendicular to the boundary of the simulation).
We refer the reader to \citet{Behroozi2020} for a full description.

For all halos in the lightcone, we use the virial mass of each halo as the `root mass' and construct a Monte Carlo realization of the merger history using an extended Press-Schechter (EPS)-based method \citep{Lacey1993, Somerville1999a, Somerville2008}.
These semi-analytically constructed dark matter halo merger trees resolve progenitor halos down to a limiting mass of $\sim 10^{10}$ \Msun\ or 1/100th of the root halo mass, whichever is smaller, for all halos. These merger trees are shown to be qualitatively similar to the ones extracted from $N$-body simulations, and the EPS method enables us to simulate galaxies over a much larger dynamic range than if we had used the merger trees extracted from the $N$-body simulation.
We note that due to the limitations of the EPS algorithm, the halo merger trees do not account for environmental influences, such as assembly bias.
We are in the process of developing a merger tree algorithm that account for these effects with machine learning methods (T. Nguyen et al. in preparation) based on $N$-body cosmological simulations with extremely high mass and temporal resolution (Yung et al. in preparation).

\subsection{The semi-analytic galaxy formation model}

Using the halos and their merger histories described in the previous section as input, semi-analytic models provide detailed predictions for the star formation histories  and a wide variety of other physical properties of galaxies, which can then be forward modelled into observable properties.
The Santa Cruz SAM consists of a collection of carefully curated physical processes that are either described analytically or derived from observations and hydrodynamic simulations.
These processes include gas cooling and accretion, star formation, stellar feedback, chemical evolution, black hole growth, and AGN feedback.
We refer the reader to the schematic flow chart (fig.~1) in \citet{Yung2022} for a comprehensive illustration of the internal workflow of the Santa Cruz SAM.

As in \citet{Yung2022} and the rest of the \textit{Semi-analytic forecasts for JWST} series papers, we adopt the fiducial `\citetalias{Gnedin2011}--\citetalias{Bigiel2008}2' model, which includes a multi-phase gas partitioning recipe motivated by numerical simulation results from \citet[][denoted by \citetalias{Gnedin2011}]{Gnedin2011}. The cold gas in the galactic disc is partitioned into a neutral, ionized, and molecular component, and an observationally-motivated \molh-based star formation recipe from \citet[][denoted by \citetalias{Bigiel2008}]{Bigiel2008} is adopted, where the slope of the relation between surface densities of SFR, $\Sigma_\text{SFR}$, and molecular hydrogen, $\Sigma_\text{\molh}$, is unity at lower gas surface densities, and the slope of the relation steepens to $\Sigma_\text{SFR} \propto (\Sigma_\text{\molh}) ^2$ above a critical surface \molh\ density.
\citet{Popping2014} and \citet{Somerville2015} implemented several models for cold gas partitioning and cold gas-based or \molh-based SF relations, and showed the impact of these different modelling assumptions on galaxy properties.
The predicted cold gas properties are compared with the IllustrisTNG simulations and ALMA observations in \citet{Popping2019a}.
\citet{Yung2019} and \citet{Yung2019a} further experimented with a subset of star-formation relations and found that the fiducial model choices adopted here best reproduce the observed evolution in the galaxy population at $4 \leq z \leq 10$.

Like any physical models that utilize analytic or `sub-grid' prescriptions, the Santa Cruz SAM contains free parameters that must be calibrated to match global galaxy observations \citep[see discussion in][]{Somerville2015}. We refer the reader to \citet{Somerville2008, Somerville2015} for detailed descriptions of the full set of parameters in the model. The Santa Cruz SAMs are typically calibrated `by hand' to reproduce a set of $z\sim0$ observational constraints, including stellar-to-halo mass ratio \citep{Rodriguez-Puebla2017}, stellar mass function \citep{Bernardi2013}, $M_\text{BH}$--$M_\text{bulge}$ relation \citep{McConnell2013}, cold gas metallicity \citep{Andrews2013, Zahid2013, Peeples2014}, stellar metallicity \citep{Gallazzi2005}, and cold gas fraction \citep{Boselli2014, Peeples2014, Calette2018}. We do not tune the models to match $z>0$ observations.
In \citet{Yung2019}, we updated the cosmological parameters to be consistent with more recent constraints from Planck, and this necessitated a minor re-calibration of the model parameters to retain agreement with the calibration observations.
The set of parameters tuned in this process are supernova (SN) feedback efficiency $\epsilon_\text{SN}$, SN feedback slope $\alpha_\text{rh}$,  star formation timescale normalization $\tau_{*,0}$, chemical yield $y$, and radio mode AGN feedback $\kappa_\text{AGN}$.
\citet{Yung2019} showed that $\epsilon_\text{SN}$ ($\tau_{*,0}$) has a significant effect on the faint (bright) galaxy populations at $z > 4$. The impact of varying $\alpha_\text{rh}$ and $\tau_{*,0}$ on the predicted galaxy populations at $4 < z < 10$ was explored in \citet{Yung2019, Yung2019a}.  On the other hand, it was shown that AGN feedback has no noticeable effect on the galaxy populations across the full range of mass and  luminosity at $z>4$.

The performance of this model configuration has been tested against observations at low redshifts \citep[$0\lesssim z \lesssim 6$;][]{Somerville2015, Somerville2021} and at high redshifts \citep[$4\lesssim z \lesssim 10$;][]{Yung2019, Yung2019a}. It has been shown that these physical models can reproduce the observed distribution functions for stellar mass and star formation rate up to $z\sim8$, as well as rest-frame UV luminosity functions up to $z\sim10$. \citet{Yung2019a} has also shown that the predicted stellar-to-halo mass ratio and other scaling relations are in good agreement with other empirical models and hydrodynamic simulations.

We note that the same satellite position re-assignment detailed in \citet{Yung2022} is also applied to the lightcones presented in this work, where the satellite positions are assigned assuming an NFW profile \citep*{Navarro1997}.
This improves the agreement of the distributions of radial distances between satellite galaxies and their central galaxies with $N$-body simulations, and is important to produce the `one-halo' term in the predicted two-point correlation functions. This is illustrated and further discussed in Section \ref{sec:clustering} and is compared to abundance matching model results from \textsc{UniverseMachine} in Appendix \ref{appendix:c}.

\begin{figure*}
    \includegraphics[width=2\columnwidth]{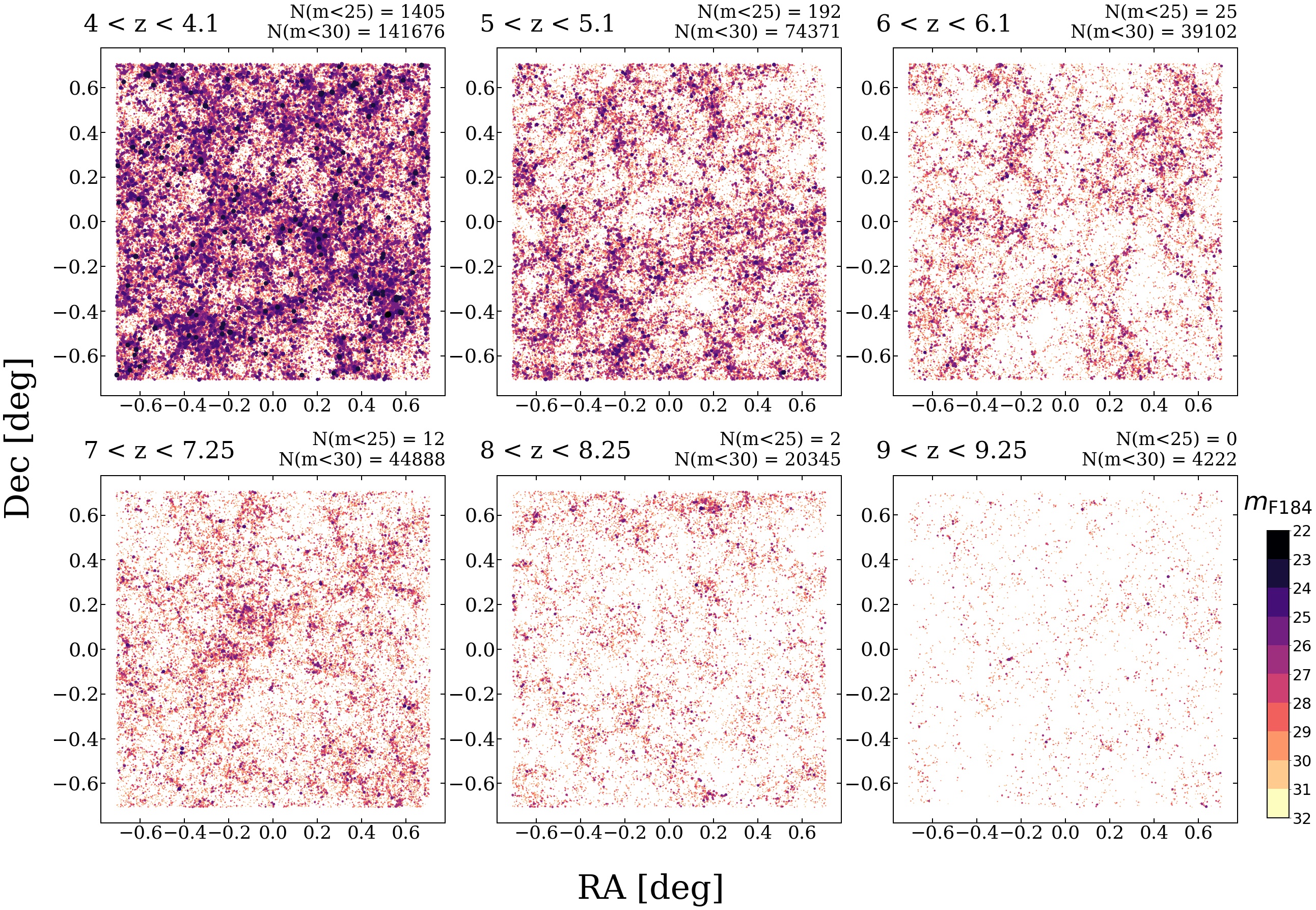}
    \caption{
        A summary of the footprint and galaxy populations in the first realization of a \twosqdeg\ lightcone at various redshift slices between $z\sim 4$ to 10. The data points are colour-coded by the observed-frame IR magnitude in the \emph{Roman} WFI F184 band. The sizes of the data points are also scaled to emphasize brighter objects and do not reflect their predicted angular sizes. In addition, the number of bright and faint objects ($m_\text{F184} < 25$ and $m_\text{F184} < 30$, respectively) within each slice is indicated at the top right corner of each panel. The specifications of this lightcone are summarized in Table~\ref{table:lightcone_specs}.
    }
    \label{fig:footprint_Roman_IR}
\end{figure*}

\begin{figure*}
    \includegraphics[width=2\columnwidth]{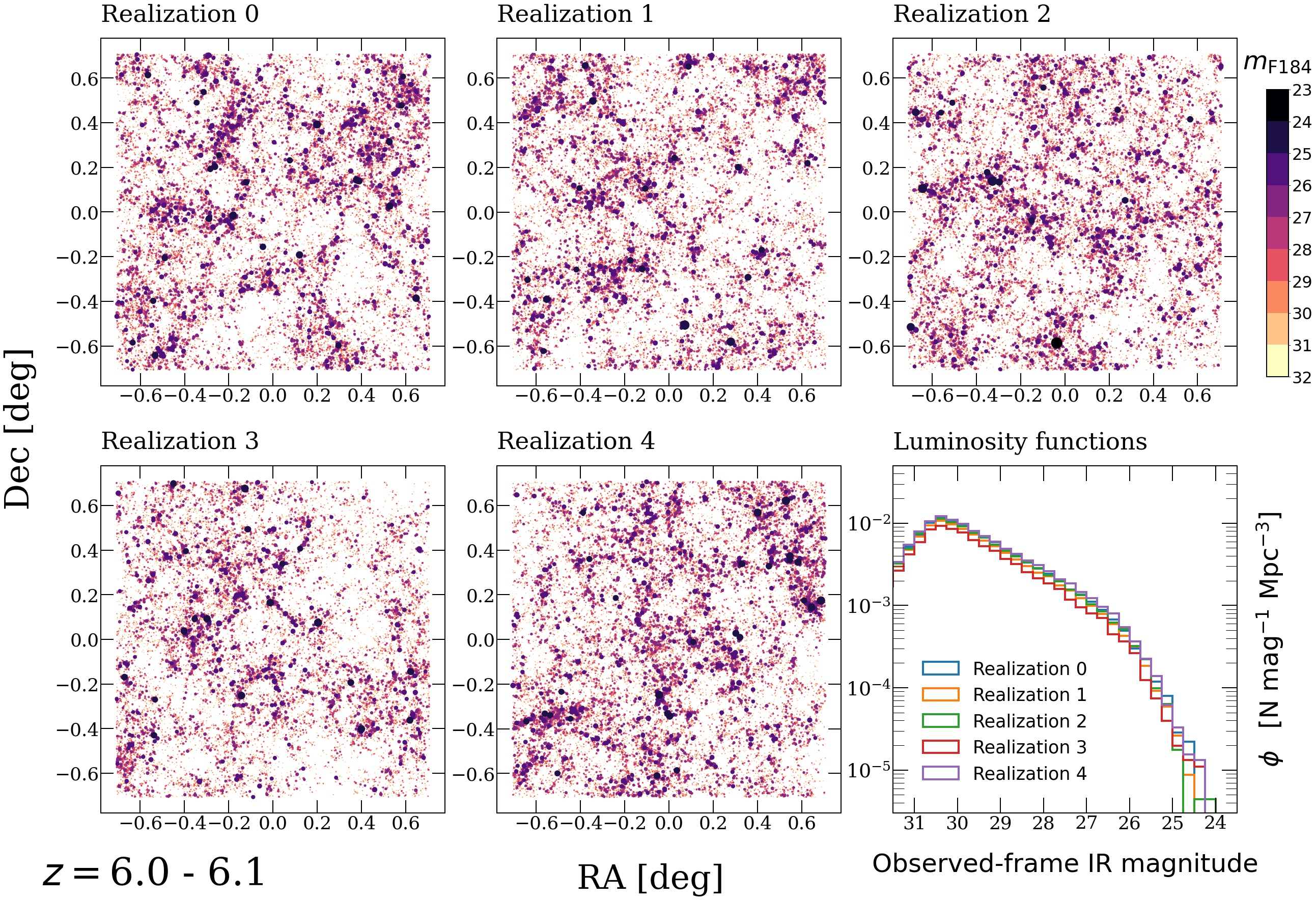}
    \caption{
        A side-by-side comparison of thin slices between $6.0 \leq z \leq 6.1$ from each of the five realizations of the \twosqdeg\ lightcones. The predicted galaxies are both size- and colour-coded by their rest-frame UV luminosity, where darker coloured, larger data points represent brighter galaxies. The size of the data point do not reflect their predicted angular sizes. The lower-right panel shows the predicted observed-frame IR luminosity functions for $6.0 \leq z \leq 6.1$ across the five realizations.
    }
    \label{fig:field_to_field_compare}
\end{figure*}

\subsection{Observables for \textit{Roman} and other facilities}

Based on the predicted star formation and chemical enrichment histories (SFHs, stored mass in bins of stellar age and metallicity), galaxies are assigned spectral energy distributions (SEDs) generated based on the stellar population synthesis (SPS) model of \citet{Bruzual2003}.
The rest-frame SEDs are used to calculate rest-frame luminosities in filter bands as presented in the mock catalogue.
In addition, quantities labelled with dust are calculated  accounting  for  the  effect  of  dust  in  the  ISM.  We assume  the  dust  attenuation  curve of \citet{Calzetti2000}. The $V$-band dust attenuation is calculated based on the surface density of cold gas and metallicity, based on a ‘slab’ model as described in \citet{Somerville2012}, but adopting the latest recalibration of ISM dust optical depth presented in \citet{Yung2021}, where the dust optical depth, $\tau_\text{dust,0}(z)$, decreases slightly at $z \gtrsim 4$ relative to the previous calibration from \citet{Yung2019} following the implementation of an updated black hole growth model that yields slightly stronger AGN feedback. This update, consistent with \citet{Yung2022}, improves the agreement between UV LF predictions and observations at $z>6$ compared to CANDELS DR1.
We have implemented a stellar mass threshold of $\log(M_{*,\text{thres}}/\text{M}_\odot) = 7$, below which we do not generate an SED, as these galaxies would likely not be observable in the wide-field surveys of interest here. This threshold is set with the mass resolution of the underlying dark matter-only simulation and the detection limit of \emph{Roman} taken into account. We note that in these mock catalogues, we do not include the contribution from nebular line or continuum emission, but plan to do so in future work (Yung, Hirschmann, Somerville et al. in prep).

The rest-frame SEDs are then redshifted according to their redshift in the lightcone,  and  observed-frame  magnitudes are computed, accounting for attenuation effects from the intervening IGM \citep{Madau1996}.
In additional to the larger simulated volume of these new lightcones, a new aspect of the  lightcone catalogues described in this work is the additional photometry from \emph{Roman} WFI\footnote{\url{https://roman.gsfc.nasa.gov/science/Roman_Reference_Information.html} (filter transmission dated Jun 14th, 2021, accessed on Nov 1st, 2021)}, \emph{Euclid}\footnote{\url{https://euclid.esac.esa.int/msp/refdata/nisp/NISP-PHOTO-PASSBANDS-V1}, access to filters granted via private communication} visible imager \citep[VIS,][]{EuclidCollaboration2022a} and Near Infrared Spectrometer and Photometer \citep[NISP-P,][]{EuclidCollaboration2022}, and \emph{Rubin Observatory}\footnote{\url{https://github.com/lsst/throughputs/tree/master/baseline}, v1.7, accessed on May 28th, 2021} \citep{Ivezic2019}.

Photometry from the large collection of filters from \emph{Hubble}, \emph{Spitzer}, and other ground- and space-based instruments as presented in the CANDELS lightcones \citep{Somerville2021} and the NIRCam broad- and medium-band photometry presented in the \emph{JWST} lightcones \citep{Yung2022} are also included in these new lightcones. Studies utilising the predicted photometry in these bands should also reference these works.
In addition, in this work, we also added photometry for instruments utilized in the SHELA survey, including DECam, NEWFIRM $K$-band, and VISTA.
Having a large collection of predictions for photometric bands from existing instruments has been shown to be useful for characterizing foreground contaminants in wide-field surveys and in the search for extreme redshift galaxies (\citealt{Leung2022}, Bagley et al. in preparation).

Rest-frame luminosities in the mock catalogues are indicated with `\texttt{\_rest}'.
Luminosities without such labels are in the observed frame.
Similarly, the luminosities of the bulge component alone are labelled with `\texttt{\_bulge}'. See Table~\ref{table:catalogue_summary} in Appendix \ref{appendix:a} for a complete list of all physical properties and photometric bands available in the mock catalogues.

\begin{table}
    \centering
    \caption{This table summarizes the dimension, area, and key configurations for the lightcones release with this work.}
    \label{table:lightcone_specs}
    \begin{tabular}{lccc}
        \hline
        Specification                & \twosqdeg\ lightcones \\
        \hline
        Dimension (arcmin)           & $84.85\times84.85$    \\
        Area (arcmin$^2$)            & 7200                  \\
        \hline
        Base simulation              & MultiDark - SMDPL     \\
        $\log M_\text{h,res}$/\Msun  & 10.00                 \\
        \hline
        $\log M_\text{*,lim}$/\Msun  & 7.00                  \\
        $M_\text{UV}$ range          &  $-16$ to $-25$       \\
        redshift range               & $0<z\lesssim10$       \\
        \hline
    \end{tabular}
\end{table}

\section{Simulated data products}
\label{sec:data_release}

Based on the physical models that have been extensively tested and, in previous works, shown to reproduce existing observations up to $z\sim10$, we present predictions for five independently sampled \twosqdeg\ fields that spans $0 < z \lesssim 10$, providing a comprehensive compilation of photometric and physical properties of galaxies (see Table~\ref{table:catalogue_summary} for full list of available quantities).
In addition, full high-resolution spectra and star formation histories are also available. These data products are useful for a wide variety of post-processing applications, such as implementing an alternative SPS model or computing photometry for additional filters.

Each of the \twosqdeg\ lightcones contains $\sim 12$ million galaxies with $\log(M_{*}/\text{M}_\odot) \geq 7$, among which $\sim 5$ million are in the rest-frame luminosity range $-16 \lesssim M_\text{UV} \lesssim -25$ or observed-frame magnitude $31 \lesssim m_\text{F184} \lesssim 22$ (both at $z\sim4$).
We note that all predicted observed- and rest-frame magnitudes presented in this work include dust attenuation, unless specified otherwise.
The key specifications of these lightcones are summarized in Table~\ref{table:lightcone_specs}.
These \twosqdeg\ lightcone have the same mass resolution as the wide-field lightcones presented in \citet{Yung2022}. However, the approximately seven times larger footprint increases the chance of including galaxies forming in more massive halos, and therefore results in better sampling at the bright end.

Fig.~\ref{fig:footprint_Roman_IR} shows the predicted star-forming galaxies from the first realization of the simulated lightcones in several redshift slices: $4.0 < z < 4.1$, $5.0 < z < 5.1$, $6.0 < z < 6.1$, $7 < z < 7.25$, $8 < z < 8.25$, and $9 < z < 9.25$.
The data points are colour-coded by their rest-frame UV luminosities.
While the sizes of the data points in this figure are scaled with their observed-frame IR luminosities in the WFI F184 filter, $m_\text{F184}$, to emphasize the bright objects, we note that the sizes of these data points do not reflect the galaxies' predicted angular sizes.
In addition, we also show the counts of galaxies in the slice within each of these panels.
This figure gives an intuitive, qualitative view of the possible evolution of object number density and large-scale structure along the light of sight in a \twosqdeg\ field.

Fig.~\ref{fig:field_to_field_compare} shows a side-by-side, qualitative comparison of thin slices of the lightcone between $6.0 \leq z \leq 6.1$ from all five realizations.
Similar to the previous figure, the data points are colour- and size-coded by the observed-frame $m_\text{F184}$ of the simulated galaxies.
We note that the colour- and size-code for this figure is chosen to be slightly different from the previous figure to better highlight the galaxy populations and large-scale structure within this redshift range.
In the last panel of Fig.~\ref{fig:field_to_field_compare}, we compare the distribution functions of $m_\text{F184}$ across the five realizations.
We show that the difference in the object counts across these five realizations is very small.
The flattening of the luminosity functions shows where the galaxy population is becoming unresolved due to the limited mass resolution of the underlying halo populations.
We deliberately show a very thin slice of the lightcone to highlight the differences across the multiple lightcone realizations.
We note that the redshift slice shown here is much narrower than the typical redshift range used in studies with photometric redshifts,  typically $\Delta z\sim 1$ \citep[]{Bouwens2015, Larson2022}, due to the relatively large uncertainties in the photometric redshift estimates.
We also note that the variance across the field realizations is larger on the bright end than in the faint end as expected (luminous galaxies are more strongly clustered at all redshifts). We will further explore the impacts of survey size and limiting magnitude on field-to-field variance in Section \ref{sec:variance}.

These wide-field lightcones have been shown to provide more robust number statistics and simulated volume that are necessary for preparatory studies for large galaxy surveys (e.g. \citealt{Finkelstein2021}; \citealt{Kakos2022}; Chworowsky et al., in preparation; Hellinger et al., in preparation) and for intensity mapping (e.g. \citealt{Yang2021a}).

Given the large number of galaxies included, these lightcones are delivered in slices by redshift. The redshift slices have $\Delta z = 0.1$ between $0 \leq z \leq 1$; $\Delta z = 0.25$ between $1 \leq z \leq 4$; $\Delta z = 0.50$ between $4 \leq z \leq 6$ and $\Delta z = 1$ between $6 \leq z \leq 10$. The data products presented in this work are accessible through the interactive portal Flathub (\url{https://flathub.flatironinstitute.org/group/sam-forecasts}), which allows more flexible access and download capability for galaxy catalogues across multiple redshift slices. Users may inspect and filter the data (e.g. by magnitude and/or by physical properties) and selectively download a subset of the data as needed.
Alternatively, the mock catalogues in ASCII format can be download in full at \url{https://users.flatironinstitute.org/~rsomerville/Data_Release/SAM_lightcones/}.

\begin{figure}
    \includegraphics[width=1.0\columnwidth]{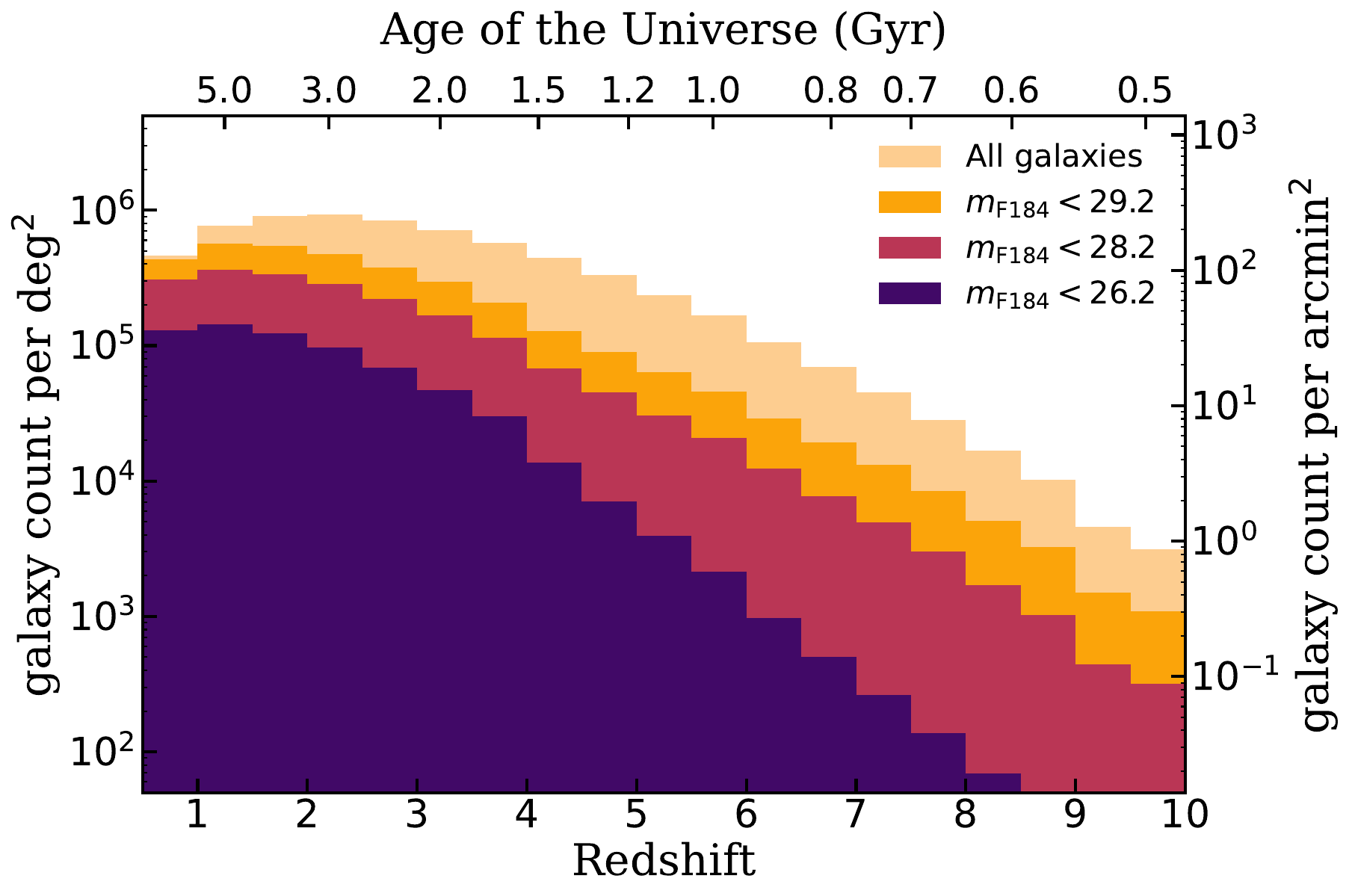}
    \caption{
        This histogram shows the (non-cumulative) number of galaxies expected normalized to unit \sqdeg\ \textit{(left axis)} and unit \sqarcmin\ \textit{(right axis)} between $0.5 < z < 10$ in bins of $\Delta z = 0.5$ for galaxies in all five realizations of \twosqdeg\ lightcones, spanning a total mock survey area of 10\;\sqdeg. On the top axis, we show the age of the Universe corresponding to the redshift indicated on the bottom axis. We show the number of objects above several survey limits expected for the high-latitude survey, moderate depth galaxy survey, and ultra-deep survey. See Table~\ref{table:survey_limits} and text for details about the survey configurations.
    }
    \label{fig:object_count_roman}
\end{figure}

\begin{table}
    \centering
    \caption{This table summarizes the $5\sigma$ detection limits, exposure time, and survey areas for anticipated \emph{Roman} surveys. The \emph{Roman} detection limits  assume the use of the WFI F184 filter. See text for full description.}
    \label{table:survey_limits}
    \begin{tabular}{llll}
        \hline
        survey type          & $m_{\text{lim},5\sigma}$ & exposure & survey area    \\
        \hline
        high-latitude survey & $26.2$   & $146$ sec& $1700$\,\sqdeg    \\
        moderate depth       & $28.2$   & $5$ hr   & $2.5$\,\sqdeg     \\
        ultra-deep survey    & $29.2$   & $20$ hr  & $0.5$\,\sqdeg     \\
        \hline
    \end{tabular}
\end{table}

\section{Results from wide-field lightcones}
\label{sec:results}

In this section, we present a set of quantitative, key predictions at high redshift that are derived from the set of \twosqdeg\ lightcones. We show the evolution of object counts (per survey area) as a function of redshift and field-to-field variance estimated for a range of survey areas.
These results are selected specifically to demonstrate the advantages of the large area coverage of these simulated lightcones.

\subsection{Evolution of galaxy demographics across redshift}
\label{sec:evolution}
In this subsection, we show predictions for the redshift evolution of volume-averaged galaxy counts and the cosmic SFR density.
Using galaxies in all five realizations of the \twosqdeg\ lightcones and the observed-frame IR magnitude in the WFI F184 filter, $m_\text{F184}$, as an example, in Fig.~\ref{fig:object_count_roman} we give an overview of the number of galaxies expected per \sqdeg\ and per \sqarcmin\ surveyed as a function of redshift.
This histogram is made with galaxies in a combined simulated area spanning 10\,\sqdeg, which contains a total of $\sim67$ million galaxies (centrals and satellites), among which $\sim34$ million, $\sim21$ million, and $\sim7$ million galaxies have $m_\text{F184} <$ 29.2, 28.2, and 26.2, respectively.
In addition, we show the corresponding age of the universe on the top horizontal axis.

This figure can be used as a look-up table to quickly estimate the number of objects expected in a large survey. Note that this figure is based on a simulated area that is $\sim35$ times bigger than one of the wide-field \emph{JWST} lightcones considered in \citet{Yung2022}, and therefore is statistically more robust and less susceptible to field-to-field variation.
In addition, we consider three discrete magnitude limits that are representative of anticipated future \emph{Roman} surveys.
These include an extremely wide but relatively shallow high-latitude survey\footnote{\url{https://roman.gsfc.nasa.gov/high_latitude_wide_area_survey.html}} that is expected to cover $\sim1700$\,\sqdeg. We also include scenarios of a moderate depth survey comparable to approximately ten WFI fields, and an ultra-deep survey that spans approximately two WFI fields.
The magnitude limits for both scenarios are estimated assuming a total $\sim 500$ hours of imaging for each type of hypothetical survey, distributed equally across all available WFI filters.
The estimated survey depths for given exposure times are calculated based on the latest available Anticipated Performance Tables\footnote{\url{https://roman.gsfc.nasa.gov/science/apttables2021/table-exposuretimes.html}} and are summarized in Table~\ref{table:survey_limits}.
In addition to showing counts per\,\sqdeg, we also show the count per \sqarcmin\ for quick reference.
For instance, we show that a moderate-depth survey reaching $m_\text{F184} \sim 28$ would yield significantly more sources at $z>8$ than the shallower high-latitude survey per unit survey area.

\begin{figure}
    \includegraphics[width=1.0\columnwidth]{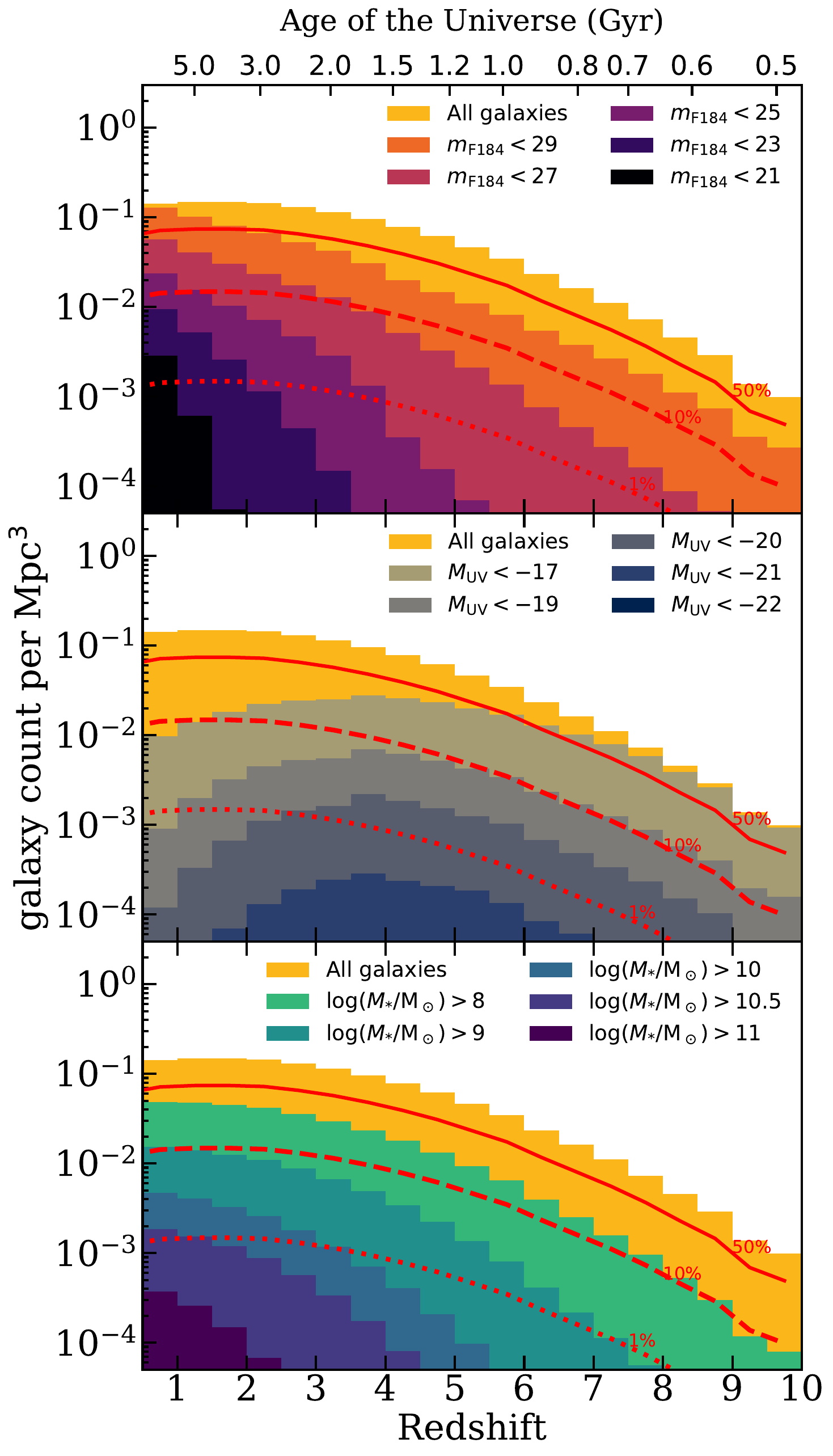}
    \caption{
        These histograms show the predicted number of galaxies normalized to unit cubic co-moving-Mpc between $0.5 < z < 10$ in bins of $\Delta z = 0.5$. The number density of all predicted galaxies is coloured in yellow. The galaxy populations are then broken down by observed-frame IR magnitude in the \emph{Roman} WFI F184 band (\textit{top}), rest-frame UV magnitude (\textit{middle}), and stellar mass (\textit{bottom}).
        In all three panels, we mark the levels of 50\%, 10\%, and 1\% of the total galaxy number counts with red solid, dashed, and dotted lines, respectively.
    }
    \label{fig:count_combo}
\end{figure}

\begin{figure}
    \includegraphics[width=1.0\columnwidth]{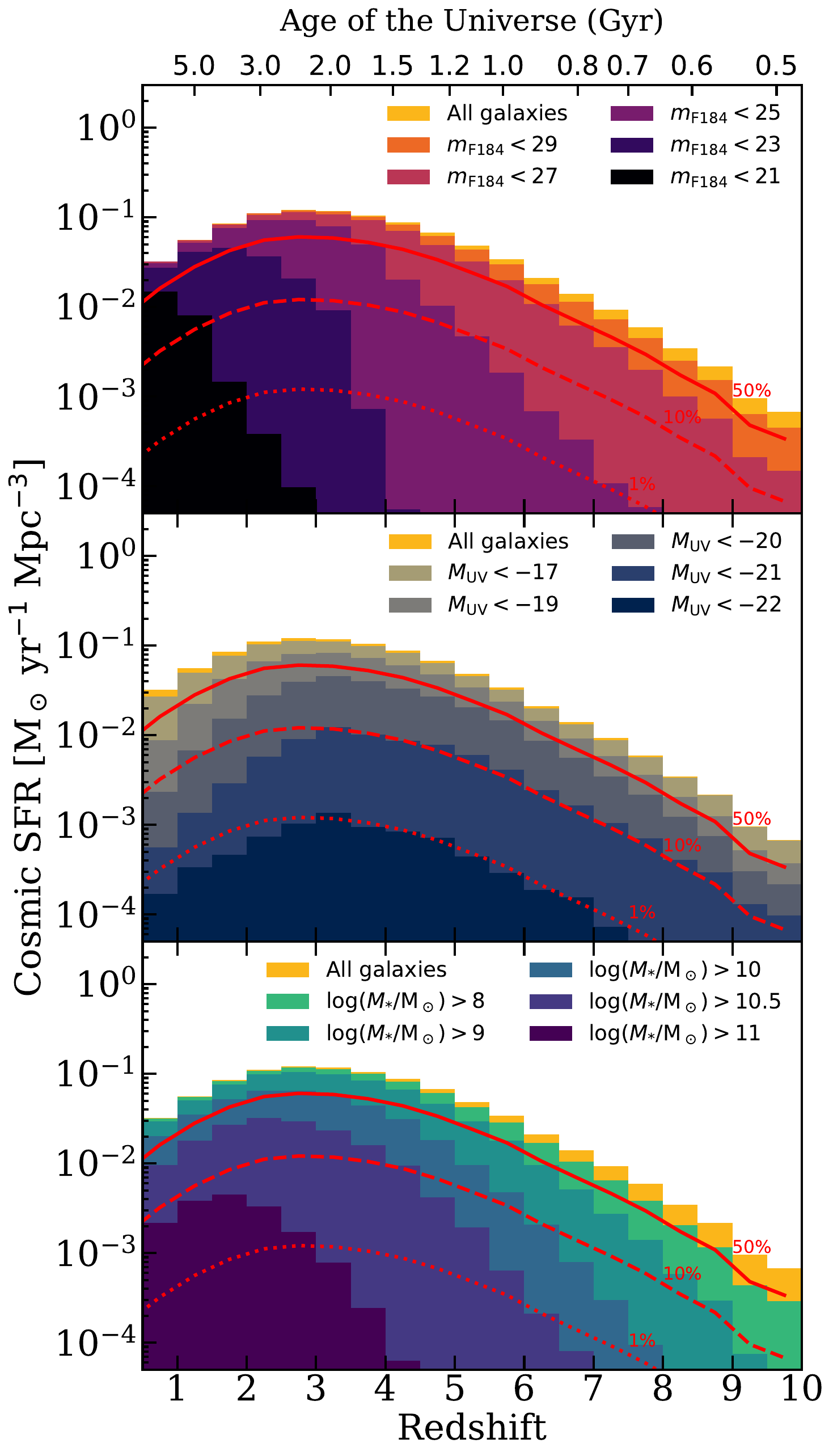}
    \caption{
        The cosmic SFR density as a function of redshift (bottom axis) or age of the Universe (top axis) between $0.5 < z < 10$ in bins of $\Delta z = 0.5$. The SFRD of all predicted galaxies is coloured in yellow. The galaxy populations are then broken down by observed-frame IR magnitude in the \emph{Roman} WFI F184 band (\textit{top}), rest-frame UV magnitude (\textit{middle}), and stellar mass (\textit{bottom}).
        In all three panels, we show lines corresponding to 50\%, 10\%, and 1\% of the total cosmic SFRD with red solid, dashed, and dotted line types, respectively.
    }
    \label{fig:sfrd_combo}
\end{figure}

The number counts of objects shown in this figure take into account several effects that impact the galaxy populations at low redshift in a degenerate manner, including the increasing dust attenuation, which preferentially affects massive galaxies. The WFI F184 filter is ideal for detecting rest-frame UV radiation for galaxies at $4 < z < 10$, but the wavelength range of the same filter begins to probe rest-frame optical and longer wavelengths for galaxies at $z \lesssim 4$. Furthermore, the number of objects \textit{per unit area} decreases as expected as the physical volume enclosed decreases. Together, these factors result in the decrease and flatting of the total number of objects expected in the F184 band at $z \lesssim 2.5$. We note that the number of objects per co-moving volume does not decrease as shown in the top panel of Fig.~\ref{fig:count_combo}.

We note that this plot provides an idealized estimate of counts of galaxies down to the stated limit. We refer the reader to Bagley et al. (in preparation) for a more detailed look into galaxy colour selection based on the mock \emph{Roman} photometry, including Lyman-break selection.

In a similar spirit, we also provide predictions for the number density of galaxies as a function of redshift.
Fig.~\ref{fig:count_combo} shows histograms of the predicted counts of galaxies per co-moving \cubeMpc\ as a function of redshift.
In the panels of this figure, the galaxy population is broken down by observed-frame IR magnitude in F184, rest-frame dust-attenuated UV luminosity, and stellar mass.
This provides an overview of the evolution in the galaxy population across cosmic time broken down by these key properties.
However, we note that the vertical axis is shown in log scale and therefore the area carved out by the histogram does not directly correspond to the proportion of their contribution. We therefore mark the 50\%, 10\%, and 1\% of the total number of galaxies available in the lightcone in all three panels.

We note that the full set of galaxy samples shown here include all galaxies available in the predicted lightcones, which contains halo populations that are complete down to $\log M_h/\text{M}_\odot = 10$.
However, the corresponding $m_\text{F184}$, $M_\text{UV}$, and $M_*$ cut-off due to the halo mass resolution limit evolves with redshift.
We refer the reader to figs.~6 and 7 in \citet{Yung2022} for where the flattening occurs in the one-point distribution functions for the observed-frame $m_\text{F184}$ and rest-frame $M_\text{UV}$. Given that these lightcones are configured with identical halo mass and stellar mass limit as the wide-field lightcones presented in that work, where the flattening occurs is expected to be the same as those presented in \citet{Yung2022}. We also refer the reader to fig.~17 in \citet{Yung2019a} for the stellar-to-halo mass relation from $z = 4$ to 10 predicted with the same model used in this work. Those results can be used to estimate the stellar mass for a given halo mass limit at a given redshift.

The top panel of this figure effectively shows the evolution of cumulative number densities of galaxies as a function of redshift at various $m_\text{F184}$ detection limits. This is equivalent to showing the redshift-evolution of the cumulative counts (number density) of galaxies with $n(<m_\text{lim})$ (also see fig.~14 in \citet{Yung2019}).
The middle and bottom panels also work the same way, which illustrate the redshift-evolution of the cumulative number density of galaxies by rest-frame UV luminosity function and stellar mass function.
This set of figure panels offers a new perspective on the predicted evolution of apparent magnitude functions, rest-frame UV luminosity functions, and stellar mass functions that are difficult to achieve with conventional one-point distribution functions.
For example, a quick comparison between the $M_\text{UV}$ \textit{(middle)} and $M_*$ \textit{(bottom)} panel shows that dust plays an important role in suppressing the rest-frame luminosity for massive galaxies, as the number of galaxies across all mass ranges increases steadily at $z \lesssim 4$ towards lower redshift, but the number of UV-bright galaxies declines.

Similarly, Fig.~\ref{fig:sfrd_combo} show the redshift evolution of cosmic SFR density (SFRD) as a function of redshift, which is a commonly used diagnostic quantity that measures the collective evolution of the integrated SFR for a galaxy population across time (see \citealt{Madau2014} for review and discussion therein).
The cosmic SFRD predicted by the Santa Cruz SAM has been shown to be in good agreement with a variety of observations integrated down to $M_\text{UV} \sim -17$ \citep{Bouwens2015, Finkelstein2015, McLeod2016, Oesch2018} and the empirical model of \citet{Behroozi2013a} (see fig.~9 in \citet{Yung2019}).
We also show a comparison of the predicted cosmic SFR with other empirical models and a subset of observational constraints in Section \ref{sec:model_comparison}.
In this figure, we break down the contribution to the cosmic SFR budget from galaxies by the $m_\text{F184}$, $M_\text{UV}$, and $M_*$ of predicted galaxies with thresholds matching the ones in Fig.~\ref{fig:count_combo}.
We also mark the 50\%, 10\%, and 1\% of the total cosmic SFR budget to guide the eye.

A side-by-side comparison of Figs.~\ref{fig:count_combo} and \ref{fig:sfrd_combo} also helps visually correlate the redshift evolution of galaxy number densities and their contributions to the cosmic SFR budget.
For instance, the bottom panels from both figures show that galaxies with $\log(M_*/\text{M}_\odot) < 8$ made up over 50\% of the galaxy populations but contribute to very little to the overall cosmic SFR.
Similarly, the middle panels show that galaxies with $M_\text{UV} < -20$ made up approximately 3\% of the population by number between $4 \lesssim z \lesssim 10$, and are responsible for $\sim 50$\% for the cosmic SFR budget.

\subsection{Field-to-field variance and survey area}
\label{sec:variance}

In this section, we leverage the combined 10\,\sqdeg\ simulated fields to conduct controlled experiments quantifying the dependence of field-to-field variance on survey area. A similar experiment was carried out in \citep{Yung2022} utilizing 40 simulated lightcones, each $\sim 1000$\,\sqarcmin. The larger contiguous areas covered by the \twosqdeg\ lightcones presented in this work enable more flexible sampling of sub-fields with different field sizes and geometry.

\begin{figure}
    \includegraphics[width=1.0\columnwidth]{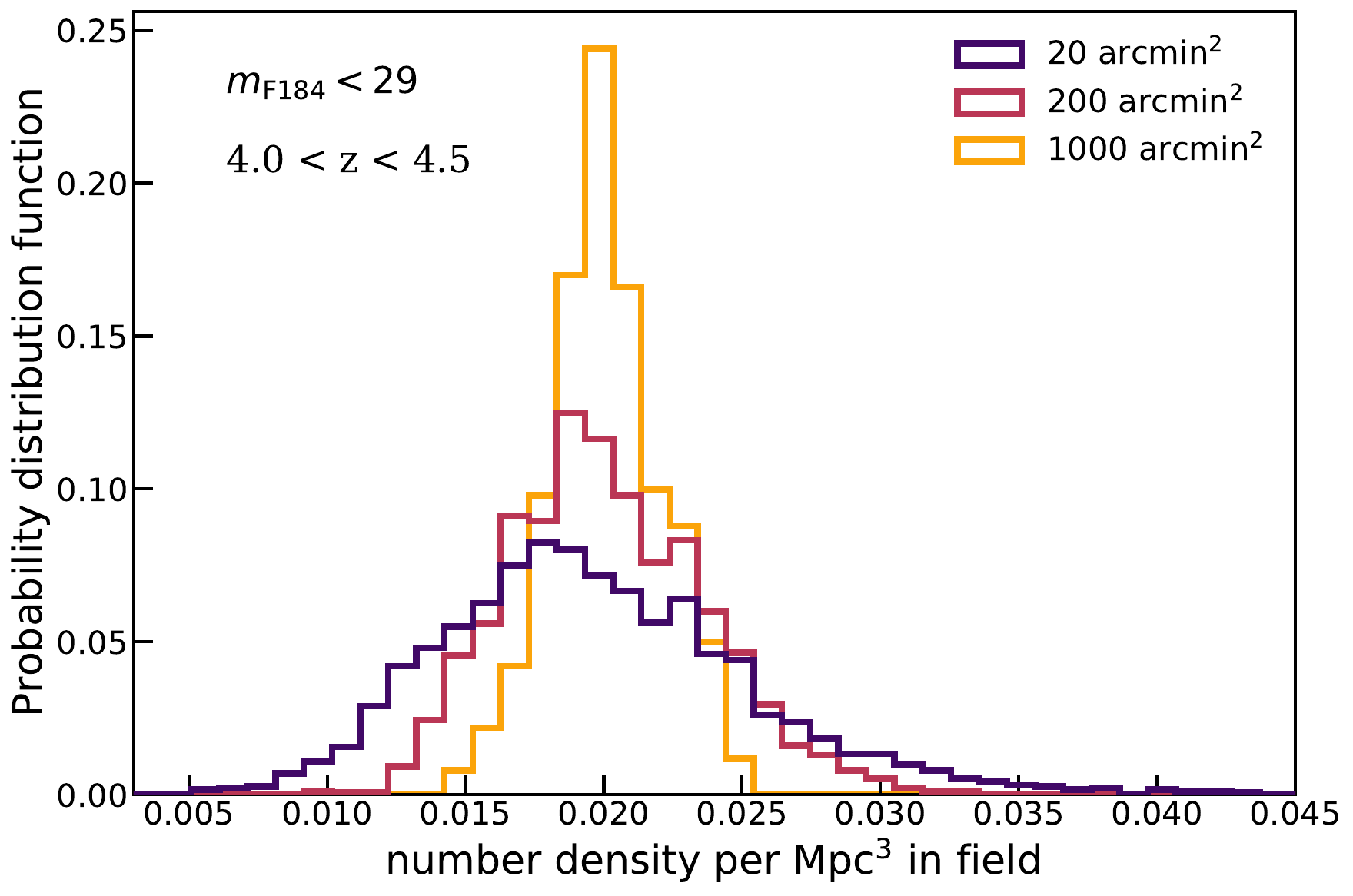}
    \caption{
        Probability distribution function for the expected number density per Mpc$^3$ for galaxies with $m_\text{F184} < 29$ between $4.0<z<4.5$ in subsampled fields of three different sizes: 20, 200, and 1000\,\sqarcmin.
    }
    \label{fig:field_count_distribution}
\end{figure}

The current generation of high-redshift surveys (with \emph{HST} and \emph{JWST}) span areas of the order of hundreds of \sqarcmin; field-to-field variation has therefore been a major source of uncertainty in the estimates of object counts.
To illustrate how future generation wide-field instruments will be able to improve this situation,
in this exercise, we consider three field sizes across multiple orders of magnitude, chosen to represent iconic deep surveys of the past and give a flavor for future surveys. We consider a case of 20\,\sqarcmin, which is comparable to the legacy HUDF or two \emph{JWST} NIRCam pointings, 200\,\sqarcmin, which is comparable to the size of a legacy CANDELS field, and 1000\,\sqarcmin, which is comparable to a single pointing of \emph{Roman} WFI.
Fig.~\ref{fig:field_count_distribution} shows the probability distribution for the number density per Mpc$^3$ for galaxies with $m_\text{F184} < 29$ between $4.0<z<4.5$ when we repeatedly subsample the \twosqdeg\ lightcones for sub-fields of the aforementioned sizes. This example demonstrates how the field-to-field variance is affected by the survey area.
We note that the distribution is skewed to the left, with a long tail towards higher number densities. This is expected in the standard theory of structure growth via gravitational collapse. Thus it is important to keep in mind that although the field-to-field \emph{variance} is often quoted, the probability distribution for over-densities can be significantly non-Gaussian.

\citet{Finkelstein2021} have utilized a similar subsampling method to investigate a possible over-density in the observed EGS field compared to the cosmic average at $z \gtrsim 9$.
A companion work will explore a similar effect in rest-frame luminosity functions (Bagley et al., in preparation).

\citet{Somerville2004} and \citet{Moster2011} provided an analytic prescription to calculate the field-to-field variance for CANDELS-sized deep pencil beam surveys for $z\sim 0.5$ to 4, based on an empirically established sub-halo abundance matching (SHAM) model for stellar mass selected galaxies. In this work, we make use of the very wide lightcones to empirically calculate the field-to-field variance for magnitude- and mass-selected galaxies at high redshift in survey fields with different sizes.
Following the steps detailed in \citet{Somerville2004} and the modification in \citet{Yung2022}, the relative cosmic variance (with shot noise removed) can be expressed as
\begin{equation}
    \sigma^2_v = \frac{\langle n^2\rangle - \langle n \rangle^2}{\langle n \rangle^2} - \frac{1}{V \langle n \rangle} \text{,}
\end{equation}
where $\langle n \rangle$ and $\langle n^2 \rangle$ denote the mean and variance of object number density $n$, respectively, and are the first and second moments of the probability distribution function $P_N(V)$ with the volume, $V$, factored out and cancelled.
In this calculation, $V$ is fixed at an assigned value such that $V\langle n \rangle > 1$ even in subregions with extremely low number density (e.g. $n < 1$ per unit volume when the number count in the field $N > 0$).

\begin{figure}
    \includegraphics[width=1.0\columnwidth]{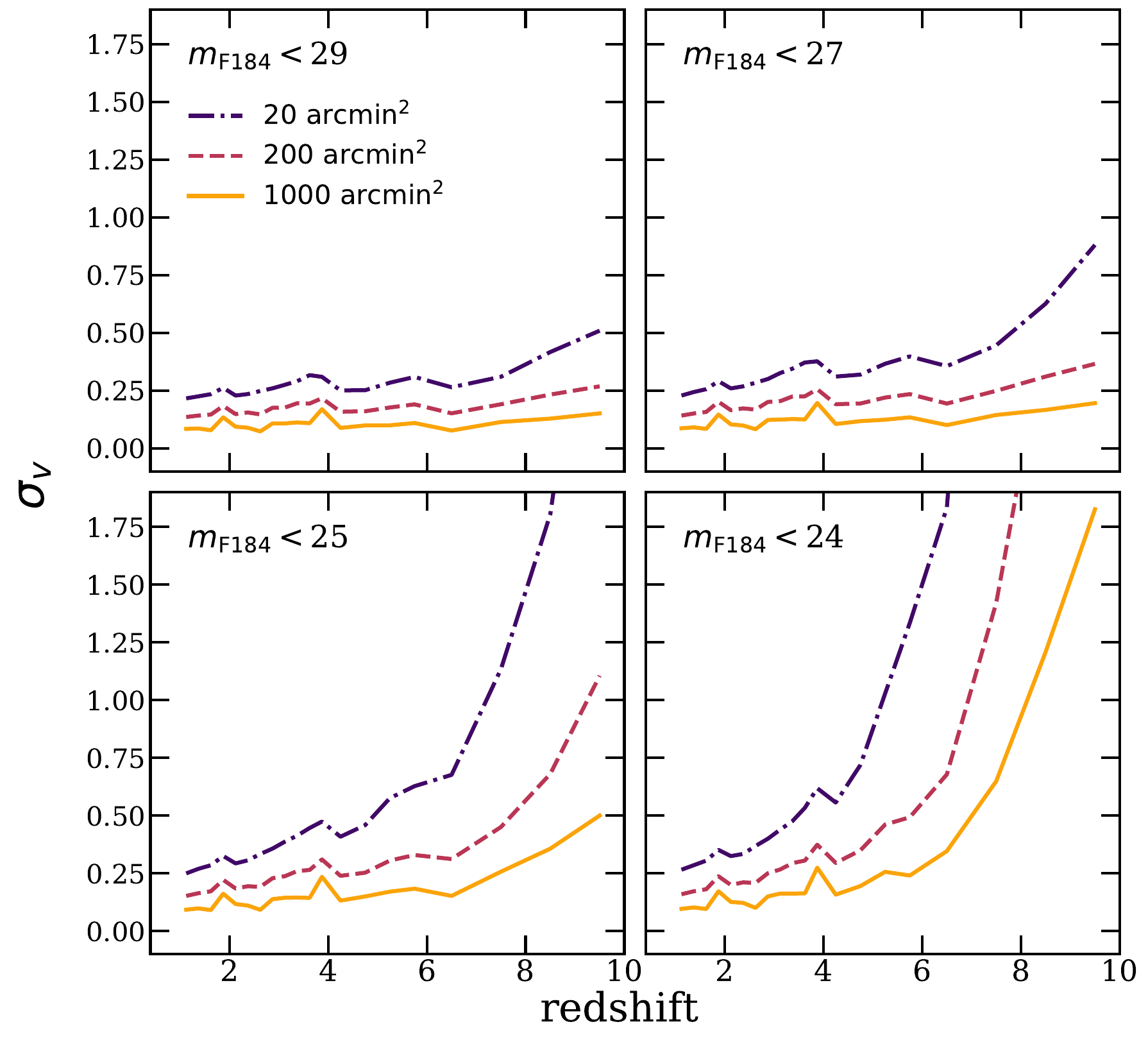}
    \caption{
        Root field-to-field variance as a function of redshift between $1 < z < 10$ for galaxies with $m_\text{F184} < 29$, 27, 25, and 24 in survey sizes that represent past and future deep galaxy surveys. We consider survey areas of 20\,\sqarcmin, 100\,\sqarcmin, and 1000\,\sqarcmin\, represented by dashed, dot-dashed, and dotted lines, respectively. See text for details regarding sampling.
    }
    \label{fig:variance_byMag}
\end{figure}

\begin{figure}
    \includegraphics[width=1.0\columnwidth]{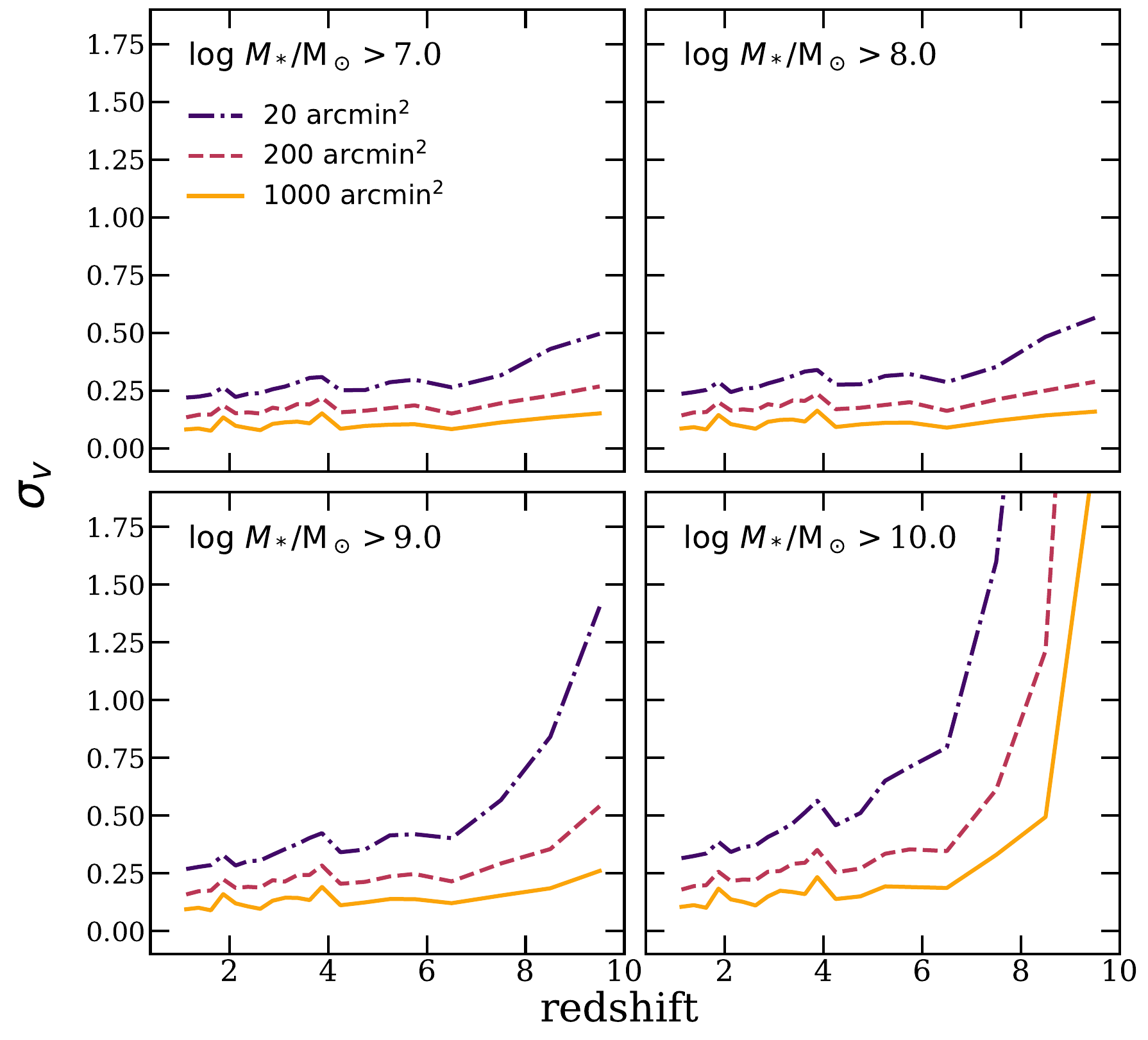}
    \caption{
        Root field-to-field variance as a function of redshift between $1 < z < 10$ for galaxies with $\log(M_*/\text{M}_\odot) > 7$, 8, 9, and 10 in survey sizes that represent past and future deep galaxy surveys. We consider survey areas of 20\,\sqarcmin, 100\,\sqarcmin, and 1000\,\sqarcmin\, represented by dashed, dot-dashed, and dotted lines, respectively. See text for details regarding sampling.
    }
    \label{fig:variance}
\end{figure}

In this exercise, we calculate the field-to-field variance empirically with predicted galaxies over the range of $1 \lesssim z \lesssim 10$ by repeatedly subsampling the lightcones with square regions and calculating the cosmic variance based on the number of galaxies (both central and satellite) found within the subregions.
We consider three distinct survey sizes: 20\,\sqarcmin\ (approximately the size of HUDF), 200\,\sqarcmin\ (on the order of current generation surveys), and 1000\,\sqarcmin\ (approximately the size of wide-field lightcones from \citet{Yung2022}). For 20\,\sqarcmin, we sample a total of a total of 3000 regions, 1000 boxes each from the first three realizations; for 200\,\sqarcmin, we sampled a total of 2500 regions, 500 boxes each from five realizations; and for 1000\,\sqarcmin, we sampled a total of 500 regions, 100 boxes each from five realizations.
These calculations share the same redshift binning as the lightcone dataset (see Section \ref{sec:data_release}).
The total volume covered by the one thousand 20 \sqarcmin\ fields is about 2.8 times larger than the total volume of the three \twosqdeg\ lightcones from which they are drawn, and for the 200 \sqarcmin\ and 1000 \sqarcmin\ fields, the ratio is 13.9. As a result, our calculation will not fully sample the density probability distribution of the larger equivalent volume. However, we have confirmed that this has a negligible effect on the estimates of the variance of the distribution, and mainly affects the tails.

\begin{figure*}
    \includegraphics[width=2.0\columnwidth]{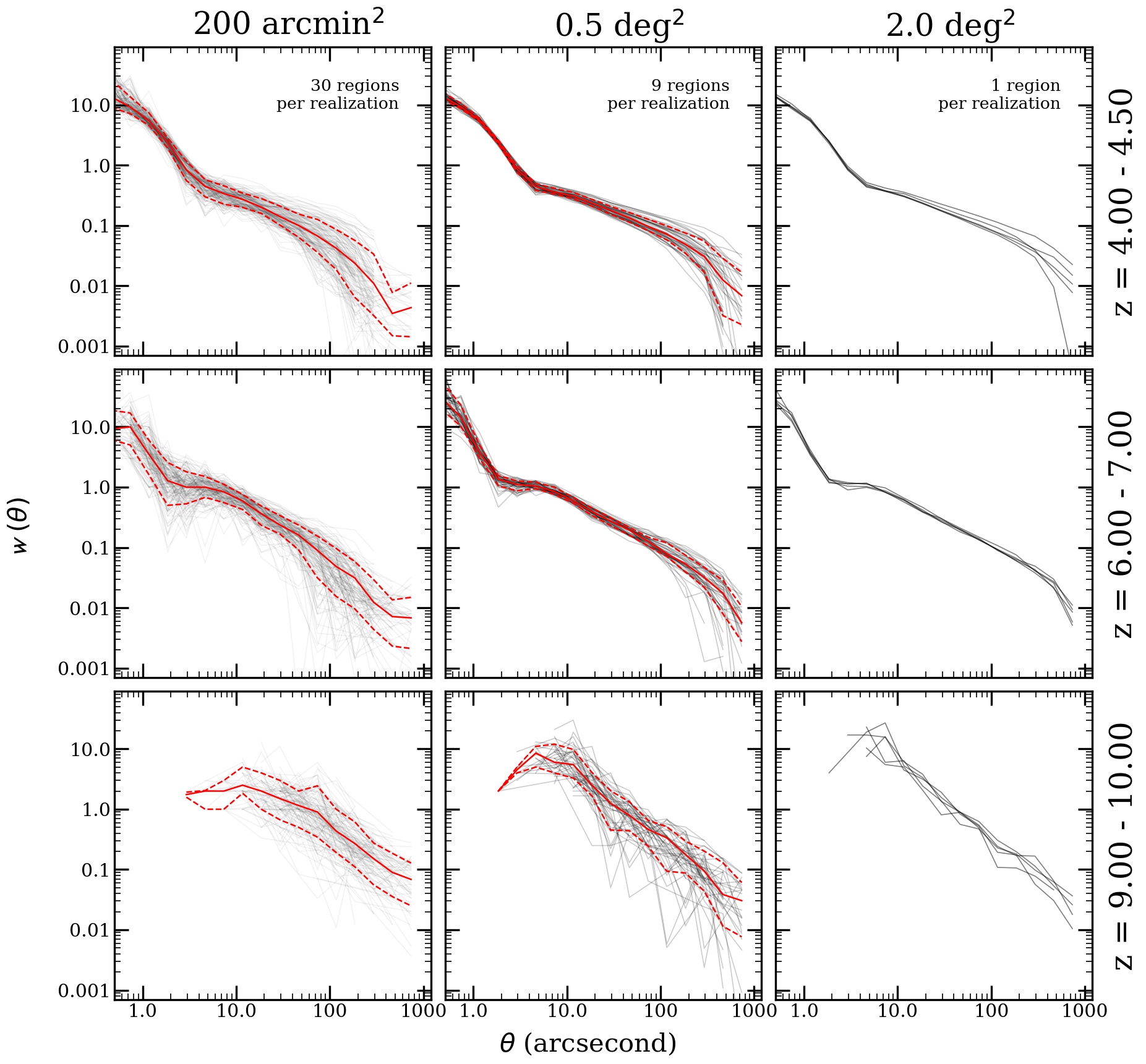}
    \caption{
        Predicted angular 2PCF, $\mathit{w}(\theta)$, calculated for galaxies with $m_\text{F184} < 28$, which is chosen to represent the depth expected to be reached by \textit{Roman} moderate depth surveys. Galaxies are subsampled from 30, 9, and 1 subregions per lightcone realization in three distinct survey areas of 200\,\sqarcmin\ (\textit{left column}), 0.5\,\sqdeg\ (\textit{middle column}), and 2.0\,\sqdeg\ (\textit{right column}), respectively. See text for details regarding the subsampling.
        The full set of 2PCF are shown individually in grey.
        This calculation is repeated for three redshift bins: $4.0 < z < 4.5$ (\textit{top row}), $6.0 < z < 7.0$ (\textit{middle row}), $9.0 < z < 10.0$ (\textit{bottom row}).
        For the 200\,\sqarcmin\ and the 0.5\,\sqdeg\ cases, the red solid line shows the median of the sampled correlation functions, and the dashed lines mark the 16th and 84th percentiles to characterize the spread among the total of 150 and 45 individually calculated 2PCFs, respectively.
    }
    \label{fig:clustering_experiment}
\end{figure*}

We show the calculated field-to-field variance as a function of redshift for observed magnitude-selected and stellar-mass-selected galaxies in Figs.~\ref{fig:variance_byMag} and \ref{fig:variance}, respectively.
In Fig.~\ref{fig:variance_byMag}, we show the evolution of the expected cosmic variance as a function of redshift for galaxies above certain observed-frame $m_\text{F184}$ thresholds at $m_\text{F184,lim} =$ 29, 27, 25, and 24.
It is well known that variance in HUDF-like and even CANDELS-like fields can be significant both at high redshift and for rare luminous objects. We show that the variance decreases significantly, down to only a few per cent, with a single WFI pointing.
Similarly, in Fig.~\ref{fig:variance}, we show the evolution of cosmic variance for galaxies above several stellar mass thresholds at $\log(M_\text{*,lim}/\text{M}_\odot) = 7.0$, 8.0, 9.0, and 10.0.
We note that the two instances where $\sigma_v$ slightly drops at $z\sim4$ and $\sim6$ are caused by the increase of redshift bin size (e.g. $\Delta z = 0.25$ to 0.50 at $z\sim4$ and $\Delta z = 0.50$ to 1.00 at $z\sim6$). While we are considering the number density of objects found throughout the volume, doubling the redshift bins has effectively increased the volume sampled and decreases the variance among the object counts. While this can be handled with finer redshift bins, we keep the current binning as it is, closer to the redshift range from colour selection at high redshift.

These calculations indicate that the variance of past CANDELS surveys ($m_\text{F160W} < 26.5$, a few hundred \sqarcmin) and deep \emph{JWST} surveys ($m_\text{F200W} < 29$, a few hundred \sqarcmin) have a potential variance of $\sigma_v \sim 1$ and $\sim 0.25$ toward $z\sim10$, which is likely the onset of cosmic reionization (see \citealt{Yung2020a} and discussion therein).
The variance depends strongly on the luminosity or mass of the objects selected, and therefore it decreases with increasing survey depth as fainter, less clustered objects are probed. In addition, the variance decreases with increasing survey area as the clustering of all objects on these large scales is smaller.  Additionally, the variance tends to increase towards higher redshift because objects of a given mass are more strongly clustered (more biased) at high redshift.  While with the current set of lightcones, we are only able to explore such calculations up to $\sim 1000$\,\sqarcmin, it is clear that future \emph{Roman} surveys, with moderate depth surveys reaching $m_\text{lim} \sim 28.2$ to span multiple square degree and the high-latitude survey reaching $m_\text{lim} \sim 26.2$ spanning thousands of \sqdeg, will further suppress the variance to negligible values, as shown by the solid orange line in Figs.~\ref{fig:variance_byMag} and \ref{fig:variance}.

\begin{figure*}
    \includegraphics[width=1.85\columnwidth]{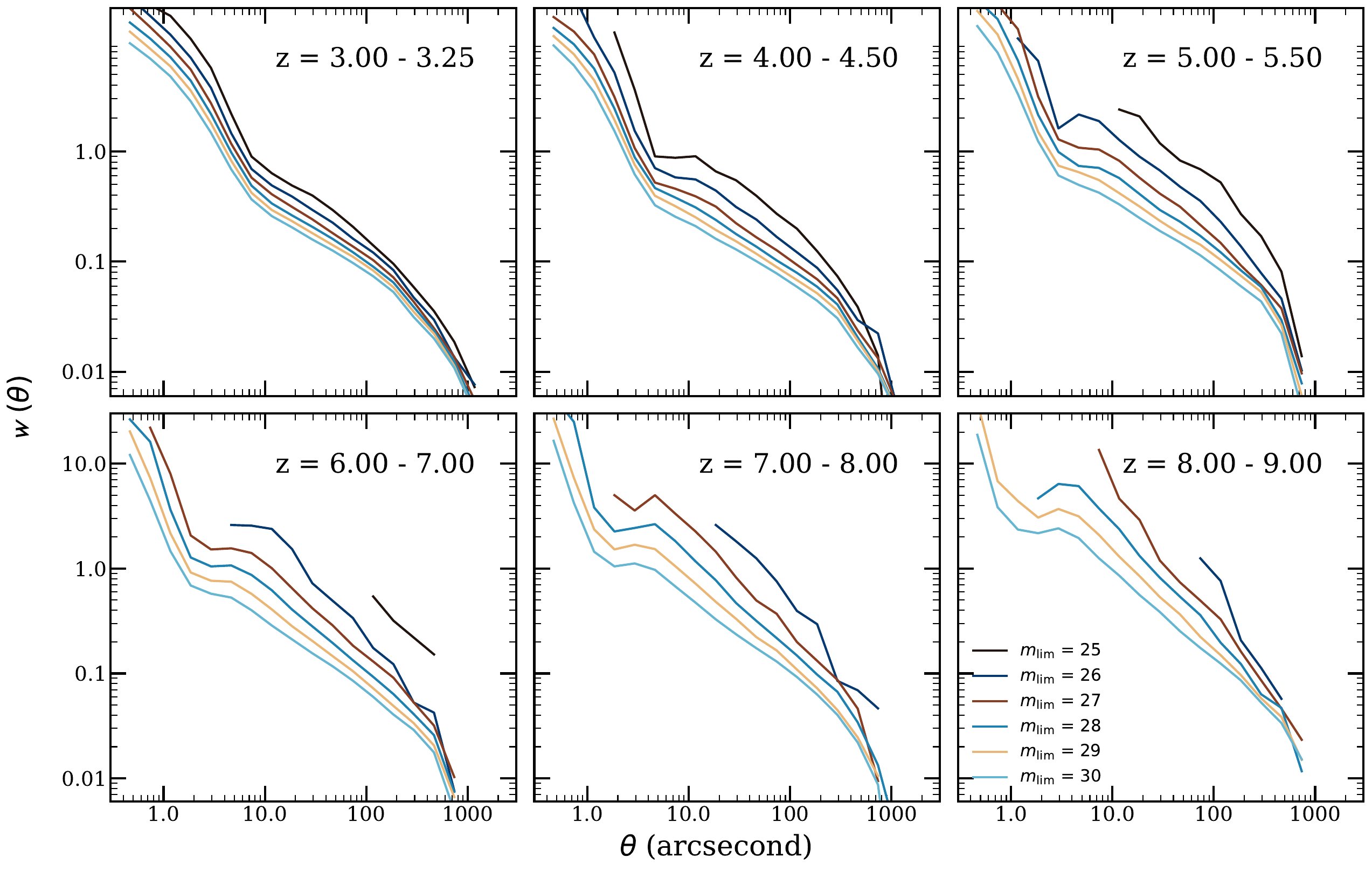}
    \caption{
        Angular correlation functions, $\mathit{w}(\theta)$, for bins across a wide range of redshifts computed for galaxy samples above several observed-IR magnitude limits. The ACFs are computed for all galaxies in each of the \twosqdeg\ lightcones, and in this figure we show the median of the five realizations.
    }
    \label{fig:clustering_experiment_mag}
\end{figure*}

\subsection{Galaxy clustering}
\label{sec:clustering}

The two-point correlation function (2PCFs) measures the excess of galaxy pair counts relative to a randomly distributed sample  \citep{Peebles1980}.
The angular auto-correlation function, $\mathit{w}(\theta)$, is a specific type of 2PCF that accounts for the projected angular separation on the sky $\theta$ between pairs of objects.
\citet{Yung2022} showed that $\mathit{w}(\theta)$ predicted from $\sim 1000$\,\sqarcmin\ lightcones constructed with the same physical model used here are in excellent agreement with observations across $1.25 \lesssim z \lesssim 7.5$ derived from the CANDELS legacy fields, \emph{Hubble} legacy deep imaging and analysis of Subaru/Hyper Suprime Cam (HSC) data presented by \citet{Harikane2016}. In addition, it was also shown that the projected 2PCF, $w_p(r_p)$, from the same lightcones are in good agreement with the PRIMUS and DEEP2 observations at $0.2 < z < 1.2$ presented by \citet{Skibba2015}.

Similar to the previous subsection, here we quantify the field-to-field variance on $\mathit{w}(\theta)$ for different survey areas.
The angular correlation functions are calculated in the same manner described in \citet{Yung2022}, utilizing
the publicly available, CPU-optimised code \textsc{corrfunc}\footnote{\url{https://github.com/manodeep/Corrfunc/}, v2.3.4} \citep{Sinha2020}.
This package adopts the Landy-Szalay estimator \citep{Landy1993}.
We note that the positions of satellite galaxies within the host halos are assigned assuming an NFW halo mass profile \citep*{Navarro1997}.
We refer the reader to section 2.1 in \citet{Yung2022} for more details.

In this section, we make use of lightcones with much larger simulated area than the ones presented in our previous work to quantify the benefits of scaling up survey areas on mitigating field-to-field variance.
Similar to the process described in Section \ref{sec:variance}, we iteratively subsample regions within the lightcones.
In this exercise, we calculate $\mathit{w}(\theta)$ for galaxies in a number of sub-fields and characterize the variance among the resulting auto-correlation functions.
We consider three distinct field sizes: 200\,\sqarcmin, which is comparable to the size of legacy CANDELS fields and current generation wide-field surveys (e.g. with \emph{JWST}); 0.5\,\sqdeg, which is approximately the size of future \emph{Roman} ultra-deep surveys (approximately 2 WFI pointings); and 2.0\,\sqdeg, which is approximately the size of future \emph{Roman} moderate-depth surveys (see Table \ref{table:survey_limits} and associated text for justification). The galaxy samples are uniformly subject to a $m_\text{F184} = 28$ detection limit, which roughly correspond to the expected depth of the moderate-depth surveys.
For the 200\,\sqarcmin\ field, we randomly sampled 30 sub-regions per simulated lightcones, totalling 150 fields.
For 0.5\,\sqdeg, we sample 9 sub-regions for each simulated lightcones, where the centres of these sub-regions are at $(x, y) \in [1/4, 2/4, 3/4]$ of the length of the axes, totalling 45 fields. This arrangement is adopted to minimize overlap among these sub-regions.
For 2\,\sqdeg, the full span of the lightcone is utilized, which gives a total of 5 fields.
In Fig.~\ref{fig:clustering_experiment}, we show ACFs computed for galaxies with $m_\text{F184} < 28$ at $4.0 < z < 4.5$, $6.0 < z < 7$, and $9.0 < z < 10$.
For field sizes of 200\,\sqarcmin\ and 0.5\sqdeg, we also mark the median and 16th and 84th percentiles with solid and dashed red lines, respectively.

This figure demonstrates the impact of variance in smaller fields, which suffer much larger uncertainties across all scale lengths.
Furthermore, the smaller fields are unable to meaningfully capture $\mathit{w}(\theta)$ at both large separation (e.g. $\theta\gtrsim500$ arcsecond), as the maximum pair separation is limited by the field size, and small separation (e.g. $\theta\lesssim20$ arcsecond), since objects at these small separations are relatively rare and only a few pairs are captured by a small field.
Similar to the cosmic variance exercise done in Section \ref{sec:variance}, the set of scenarios explored in this figure demonstrates that field to field variance in deep clustering studies at high redshift can be reduced to very low levels already with a single \emph{Roman} WFI pointing.

In Fig.~\ref{fig:clustering_experiment_mag}, using the case of the full \twosqdeg\ as an example, we explore the effects of survey depth.
We calculate the ACF individually for each realisation and show only the median among the five realizations.
For instance, the $m_\text{F184,lim} = 28$ line (shown in blue) in Fig.~\ref{fig:clustering_experiment_mag} corresponds to the right column in Fig.~\ref{fig:clustering_experiment} in matching redshift bins.
This figure provides a general overview for how the clustering statistics are affected by the survey depth.
One can also estimate the increase in scatter in the relation due to field-to-field variance by referencing results from Section \ref{sec:variance} (e.g. Fig.~\ref{fig:variance_byMag}).
As noted above, the survey depth strongly impacts the clustering of the selected galaxy population, as deeper surveys will be dominated by fainter galaxies, which are more weakly clustered. In addition, this effect does not scale uniformly across redshift, as the strength of galaxy clustering at a given mass/luminosity also depends on redshift. This is because halos of a given mass represent rarer, more clustered `peaks' at earlier times in cosmic history.
For instance, $w(\theta)$ seems to be uniformly sensitive across all scales at $z\sim3$. However, the small-scale end becomes increasingly sensitive to the survey depth towards higher redshifts.
We also note that the small separation pairs disappear for the shallow survey depths at high redshift as expected, as such bright objects at such small separations are extremely rare.

\begin{figure}
    \includegraphics[width=1\columnwidth]{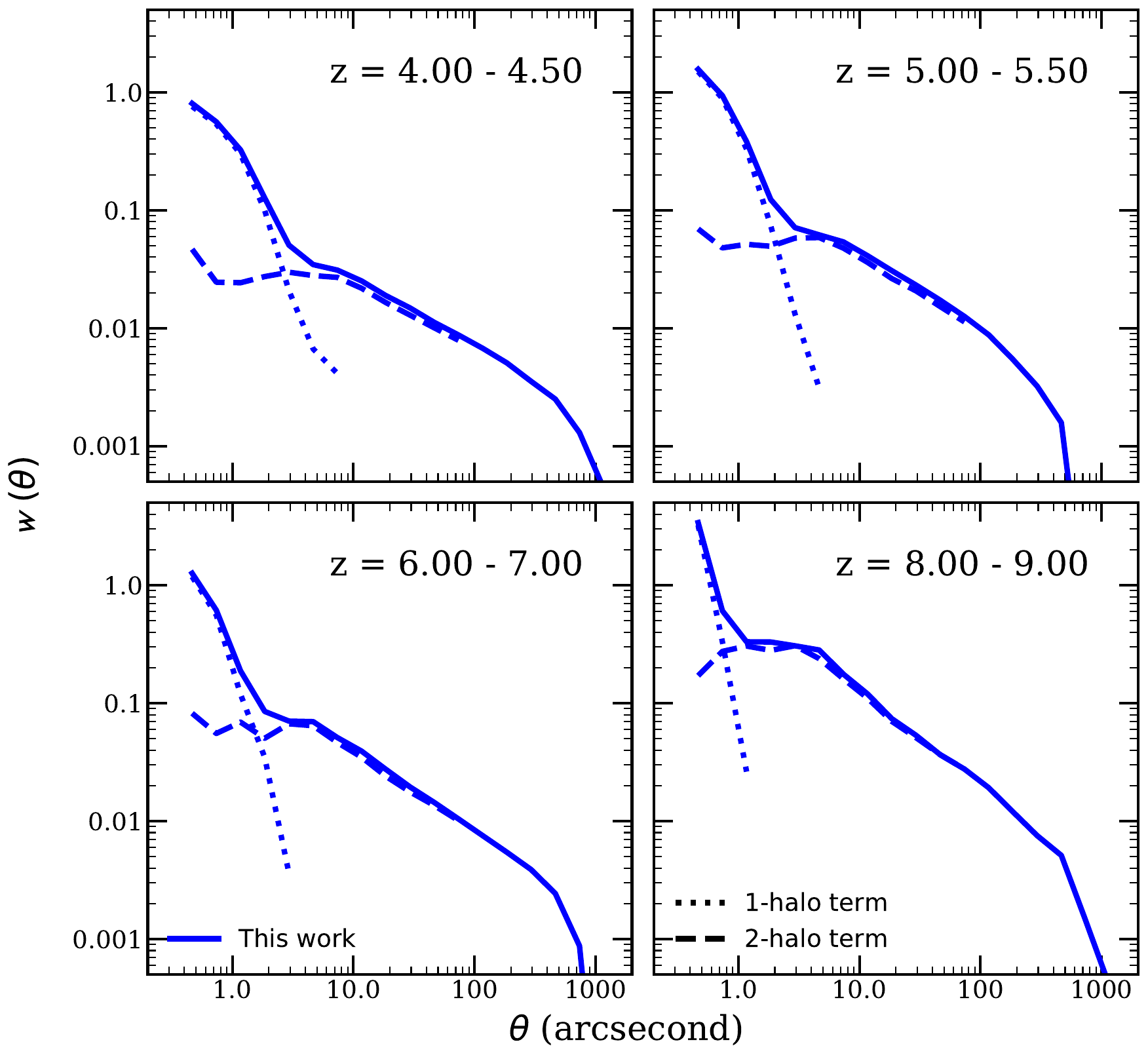}
    \caption{
        Angular correlation functions, $\mathit{w}(\theta)$, calculated for galaxies with $M_\text{UV} < -18$ in redshift bins across $z\sim 4$ to $\sim9$ from the first realization of the 2-\sqdeg\ lightcones presented in this work. In addition to the $\mathit{w}(\theta)$ calculated for the full galaxy populations (solid line), we show the one- and two-halo terms with dotted and dashed lines, respectively (see text for details).
    }
    \label{fig:clustering_models_2halo}
\end{figure}

In Fig.~\ref{fig:clustering_models_2halo}, we show the ACF calculated for galaxies in four redshift bins spanning across $4.0 \lesssim z \lesssim 9.0$ for galaxies $M_\text{UV} < -18$. In addition to the $\mathit{w}(\theta)$ calculated for the full galaxy populations, we show the one- and two-halo terms, where the `one-halo' term refers to pairs that are within the same halo and the `two-halo' term refers to pairs where each galaxy is in a distinct dark matter halo. The two-halo term is calculated only for central galaxies. The one-halo term is calculated by subtracting the two-halo term from the $\mathit{w}(\theta)$ for the full galaxy population. This figure illustrates the angular scales where we expect the one vs. two-halo term to dominate at high redshift. The angular scale where the two terms cross ranges from about 3 arcsec at $z\sim 4$ (equivalent to $\sim25$ kpc) to $\sim 1$ arcsec at $z\sim 8$--9 (equivalent to $\sim5$ kpc).

\subsection{Comparison with other lightcones}
\label{sec:model_comparison}

We compare our predictions with those from two other studies in which similar lightcones have been constructed, with galaxy properties assigned via empirical models. Here we show a direct comparison of stellar-to-halo mass ratio (SHMR), galaxy counts per \cubeMpc, and the ACFs across these models. Both simulations adopted cosmological parameters that are similar to the ones adopted in this work. In addition, halo catalogues underlying all three simulations are generated using \textsc{Rockstar} and \textsc{Consistent Trees} \citep{Behroozi2013b, Behroozi2013c}.
These simulations also all adopt a \citet{Bryan1998} virial mass definition, a \citet{Chabrier2003a} IMF, and a \citet{Calzetti2000} attenuation curve.

\textsc{UniverseMachine} is an empirical model that is optimized to reproduce a wide variety of observational constraints, including stellar mass functions, UV luminosity functions, cosmic star formation rate, specific star formation rate, etc  \citep{Behroozi2019, Behroozi2020}.
In this empirical model, the star formation is modelled as a function of
$v_{\text{M}_\text{peak}}$, $z$, and $\Delta v_\text{max}$,
where $v_{\text{M}_\text{peak}}$ is the circular velocity of the halo at the redshift where it reached its peak mass and $\Delta v_\text{max}$ corresponds to the relative change in $v_\text{max}$ over the past dynamical time.
The SFR as a function of $v_\text{max}$ and quenched fraction $f_\text{quenched}(v_\text{max})$ are iteratively tuned using Markov Chain Monte Carlo until the modelled galaxy populations matches a range of observational constraints, including stellar mass functions between $z = 0$ to 4, cosmic star formation rates between $ z=0 $ to 10, etc.~(see table 1 in \citet{Behroozi2019} associate text for detail).
This model has previously been used to interpret CANDELS observations and has been compared to the Santa Cruz SAM in \citet{Somerville2021}.
In this work, we compared to a custom version of \textsc{UniverseMachine} lightcone,
which is constructed with the same underlying halo populations as the ones used in this work.
Therefore, the \textsc{UniverseMachine} lightcone compared here has exactly the same footprint and halo mass resolution as the \twosqdeg\ lightcones presented in this work.
We note that the UniverseMachine lightcone used in this comparison is constructed based on the the exact same underlying set of halos.

The Deep Realistic Extragalactic Model (DREaM) Galaxy Catalogue\footnote{\url{https://www.nicoledrakos.com/dream}} is a lightcone created specifically for future deep galaxy surveys with \emph{Roman} \citep{Drakos2021}.
This $1$-\sqdeg\ lightcone is constructed based on a dark matter-only $N$-body simulation and galaxies from an empirical model \citep{Springel2005a, Williams2018}. The underlying dark matter-only $N$-body simulation has a box size of 115 $h^{-1}$ Mpc and dark matter particle mass of $1.5\times 10^7$ M$_\odot\, h^{-1}$. The stellar masses of galaxies are assigned to dark matter halos using the subhalo abundance matching (SHAM) technique and are constructed to match the observed stellar mass functions at $z \leq 4$ and the available observed UV luminosity functions.
Galaxies are proportionately but randomly assigned to be star-forming or quiescent, and depending on the galaxy type, various free parameters, such as $e$-folding time ($\tau$) and the age of the Universe at the time star formation started ($t_\text{start}$), are sampled from a `parent catalogue' constructed based on known scaling relations.
The star formation histories of these galaxies are then modelled using the `delayed tau model', given by $\text{SFR}(t) \propto t\,e^{-t/\tau}$ with $t$ being the time elapsed since $t_\text{start}$ (see \citealt{Williams2018} and references therein).
Based on the combination of fixed parameters ($M_*$ and $z$), randomly assigned galaxy type, and inferred parameters, synthetic SEDs are generated using the Flexible Stellar Population Synthesis (FSPS) package \citep{Conroy2009}.
The SFR for each galaxy is then inferred based on the synthetic stellar SED, which also gives $M_\text{UV}$ and other colour-related predictions, as well as the \emph{Roman} and \emph{JWST} photometry.
We note that the cosmic SFR presented in \citet{Drakos2021} is averaged over the last 100 Myr based on the assumed SFH.

\begin{figure}
    \includegraphics[width=1\columnwidth]{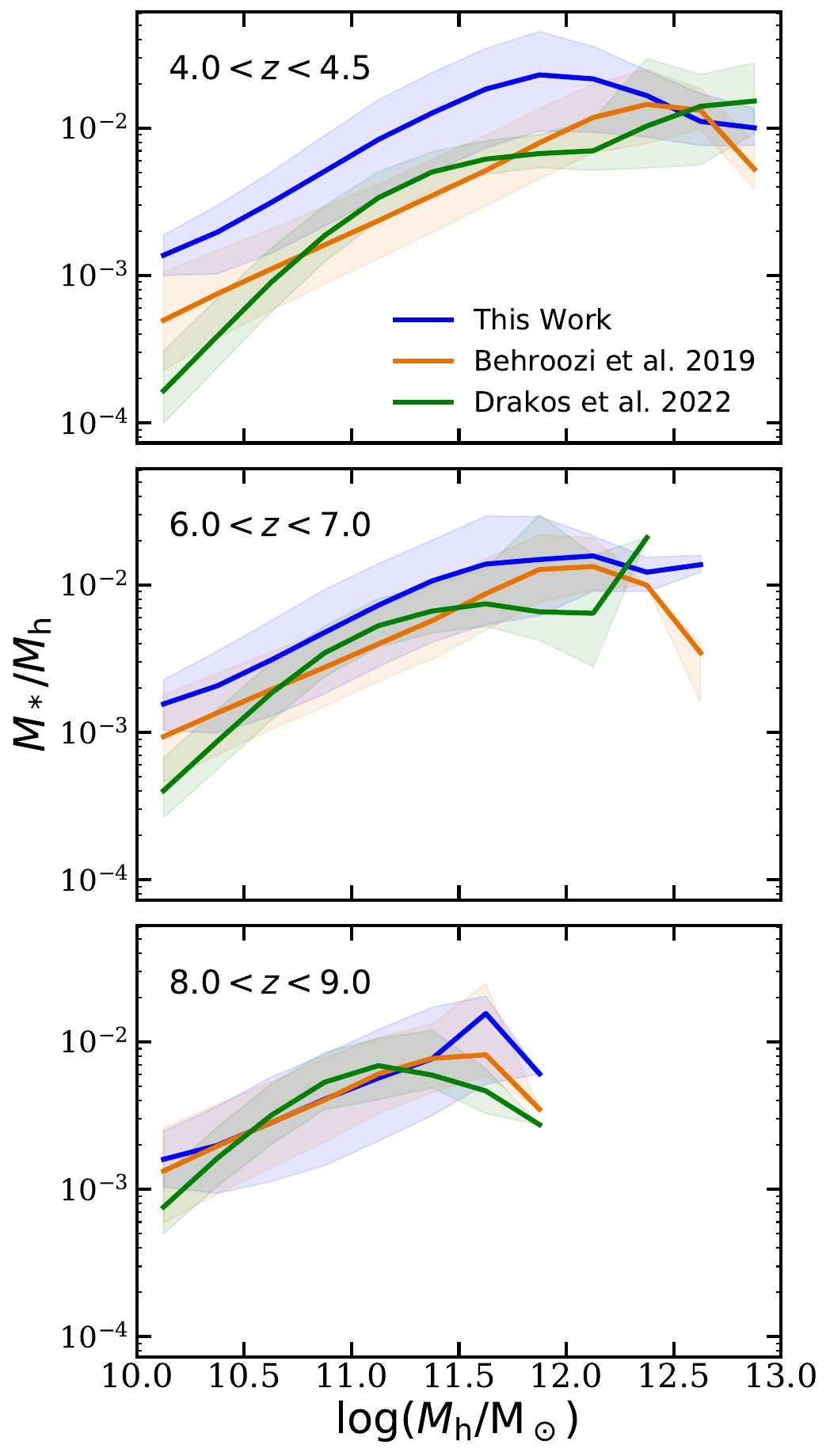}
    \caption{
        Stellar-to-halo mass ratio (SHMR) for predicted galaxies in one of the 2-\sqdeg\ lightcones (blue) between $4.0 < z < 4.5$ (\textit{top}), $6.0 < z < 7.0$ (\textit{middle}), and $8.0 < z < 9.0$ (\textit{bottom}). These results are compared to \textsc{UniverseMachine} (orange) and DREaM (green). For all models, the solid line marks the median of the relation, and the shaded region represents the 17th and 84th percentiles.}
    \label{fig:models_SHMR}
\end{figure}

\begin{figure}
    \includegraphics[width=1\columnwidth]{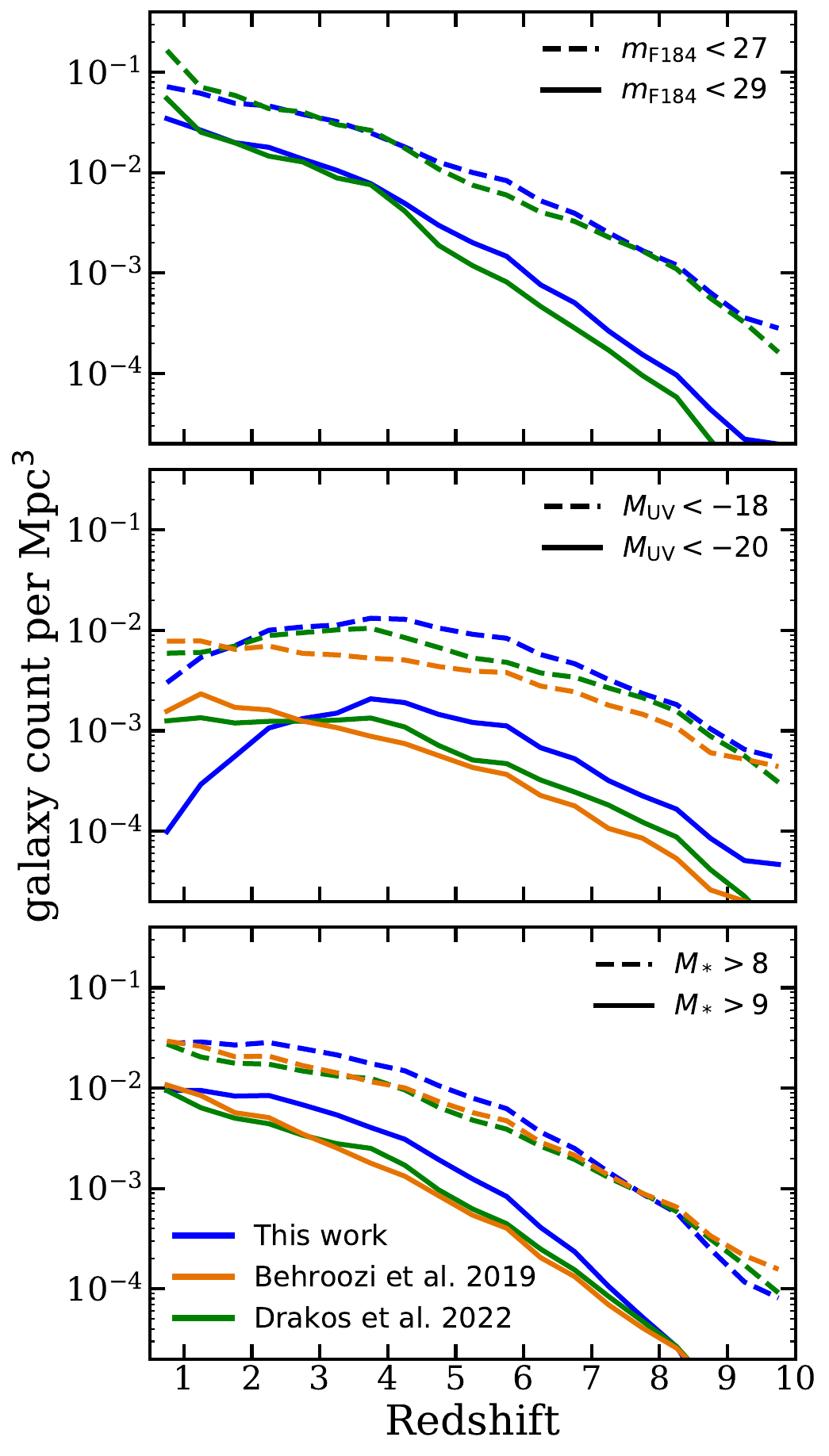}
    \caption{
        Number density of galaxies per cubic comoving-Mpc between $0.5 < z < 10$ in bins of $\Delta z= 0.5$ from one realization of the 2-\sqdeg\ lightcones presented in this work (blue), \textsc{UniverseMachine} (orange), and DREaM (green). We apply selection criteria by observed-frame IR magnitude in the \emph{Roman} WFI F184 band (\textit{top}), rest-frame UV magnitude (\textit{middle}), and stellar mass (\textit{bottom}).}
    \label{fig:count_models}
\end{figure}

\begin{figure}
    \includegraphics[width=1\columnwidth]{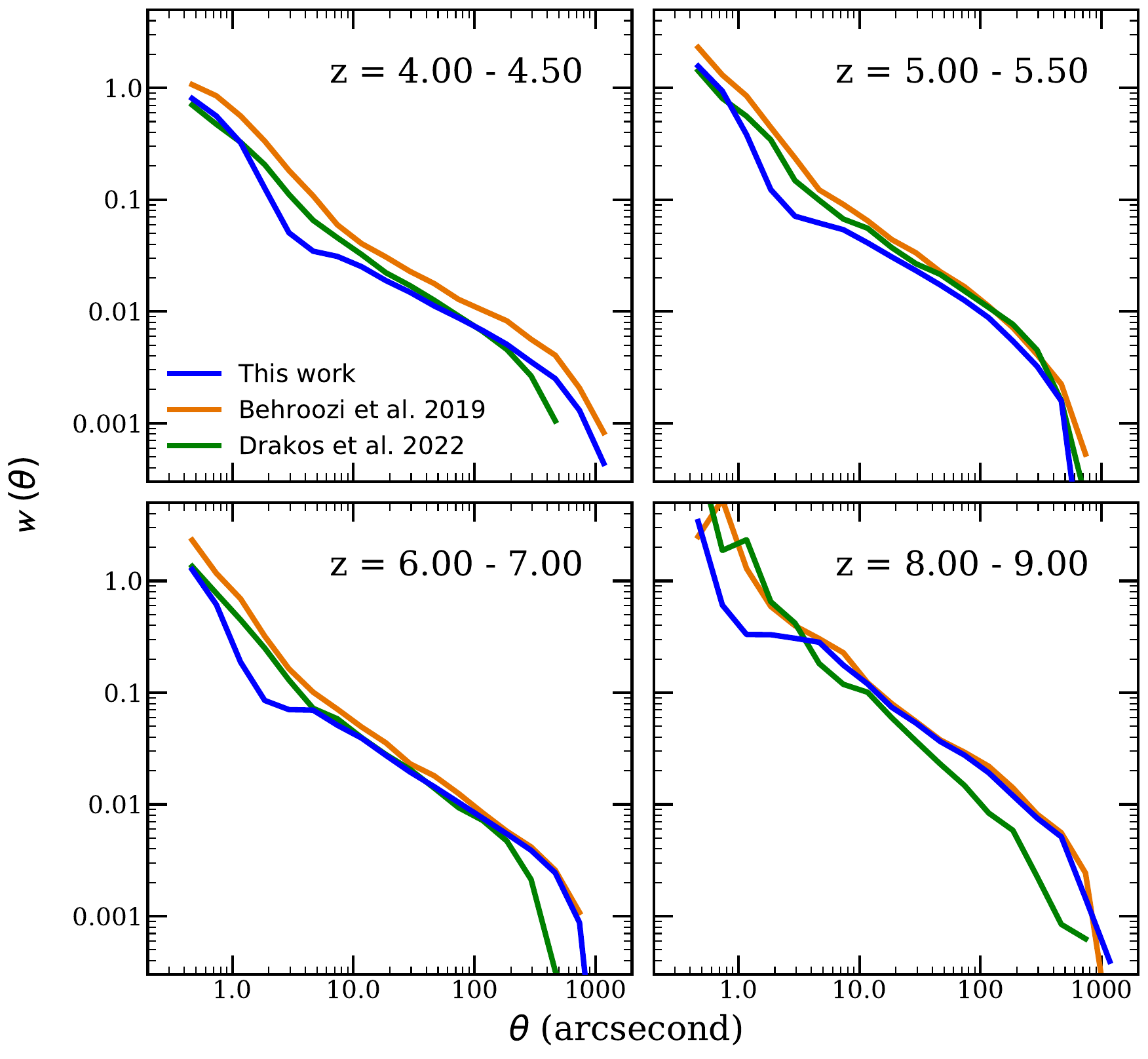}
    \caption{
        Angular correlation functions calculated for galaxies with $M_\text{UV} < -18$ in redshift bins across $z\sim 4$ to $\sim9$ from one realization of the 2-\sqdeg\ lightcones presented in this work (blue), a 2-\sqdeg\ lightcone from \textsc{UniverseMachine} (orange), and a 1-\sqdeg\ lightcone from DREaM (green).}
    \label{fig:clustering_models}
\end{figure}

It is important to note that while these three simulations share a number of basic components (e.g. cosmological parameters, dark matter-only simulations, halo finder, dust attenuation model),
which should not be a major source of differences, these three models took very different approaches to modelling SFR, $M_*$, and other galaxy properties and observables.
Fig.~\ref{fig:models_SHMR} shows the stellar-to-halo mass ratio (SHMR) for galaxies in the first realization of the \twosqdeg\ lightcones compared to galaxies from a \twosqdeg\ \textsc{UniverseMachine} lightcone and the DREaM lightcone.
This figure highlights the difference in the predicted halo occupation between the SAM and both empirical models. In particular, the SAM predicts a milder redshift evolution in SHMR compared to the empirical models.
We discuss this further in Section \ref{sec:model_comp}.
We add that this is a mass range where abundance matching, empirical models, and hydrodynamic simulations have very little consensus on even at $z\sim0$  \citep[e.g.][]{Behroozi2019, Munshi2021}.

Fig.~\ref{fig:count_models} shows the number density of galaxies per \cubeMpc\ above specific cut-off values in $m_\text{F184}$, $M_\text{UV}$, and $M_*$.
Each of the panels in this plot breaks down the predicted galaxy populations roughly into two groups.
We note that these limits are chosen to compare the number of galaxies at both a brighter (or more massive) and a fainter (or less massive) limit.
These limits are not meant to be correlated across panels, as the mass-to-luminosity ratio and the observed- and rest-frame magnitude evolve as a function of redshift.
For instance, while the SC SAM predicts a similar number of galaxies with $m_\text{F184} < 27$, this model predicts noticeably more galaxies with $m_\text{F184} < 29$ at $z > 4$.
It is not surprising to see that the number of objects as a function of rest-frame and observed-frame magnitude from these models are in good agreement, as the number density of observed IR-bright galaxies at intermediate redshift are relatively well-constrained and all of the models agree well with these observables.
However, the discrepancy in stellar mass reveals that a very different mass-to-magnitude relation is predicted by these different model approaches.
Overall, we see that the SAM predicts higher stellar mass content across $2 \lesssim z \lesssim 6$ in halos across all mass ranges at lower redshift.
\citet{Yung2019a} has shown a compilation of model predictions, including the Santa Cruz SAM, \textsc{UniverseMachine}, and \citet{Williams2018}, and show that these models predicted very different evolution of the underlying stellar mass and star formation rate. Among the models included in the comparison, the Santa Cruz SAM predicts more galaxies across a wide range of $M_*$ and SFR than the other two models at $z \lesssim 6$. We note that the SMFs predicted by these models are well within the uncertainties of the observed constraints currently available \citep[e.g.][]{Duncan2014, Song2016, Katsianis2017, Katsianis2017a}.

In Fig.~\ref{fig:clustering_models}, we show a comparison of the predicted ACFs from this work to the two empirically modelled lightcones.
This is calculated for all galaxies with $M_\text{UV} < -18$.
As shown in \citep{Yung2022}, the Santa Cruz SAM reproduces the projected 2PCF measured by PRIMUS and DEEP2 in at $0.2 \lesssim z \lesssim 1.2$ reported by \citet{Skibba2015}, at intermediate redshift $1.25 < z < 4.5$ in legacy CANDELS surveys (computed via theory catalogue presented in \citealt{Somerville2021}), and high-redshift measurement at $3.5 < z < 7.5$ reported by \citet{Harikane2016}.
The \textsc{UniverseMachine} model has been shown to reproduce the measured clustering of massive galaxies in the nearby universe up to $z \sim 0.7$ from the PRIMUS and DEEP2 survey \citep{Coil2017}. The DREaM lightcone has been shown to reproduce the $0.1 < z < 0.2$ measurements from SDSS \citep{Yang2012}.
It is interesting that, in spite of the very different modelling approaches, all three models arrive at very similar predictions for the clustering of rest-UV selected galaxies at $z \gtrsim 4$.

Differences in the ACF on the small-scale end likely arise mainly from the modelling of satellite galaxies, which is very uncertain.
The satellite galaxies in \textsc{UniverseMachine} are modelled using $\Delta v_\text{max}$ over the past dynamical time, which captures the rapid drop of $v_\text{max}$ following a major merger \citep{Behroozi2014}. $\Delta v_\text{max}$ has been shown to be a more robust measurable quantity for satellites and yield more clearly orbit- and profile-dependent satellite SFRs \citep{Onions2012}.
On the other hand, DREaM adopts the peak of the maximum circular velocity, $V_\text{peak}$, over the entire merger history as a halo mass proxy. SHAM is first performed between the SMF and halo populations, and then a random number is assigned to distinguish whether a galaxy is quiescent or star-forming. This assignment does not distinguish whether a halo is a central or sub-halo, and therefore the environmental dependence of galaxy quenching is not accounted for. In the SC SAMs, galaxies are starved of new gas cooling once they become satellites, and are therefore probably over-quenched. For that reason, our satellite galaxies are in general fainter compared to the other two models and are more susceptible to the magnitude cut applied. Additional comparison for the satellite populations in these models is available in Appendix \ref{appendix:c}.
It is not surprising that the agreement at small scales is closer between the two empirical models, as although the modelling of satellites differs in detail, it is quite similar in spirit. On the other hand, the Santa Cruz SAM seems to predict a stronger `shoulder' between the one- and two-halo terms in the ACF. Future observations should thus be able to constrain the physical processes that shape satellite galaxy properties.
The differences on the very large-scale end of the ACFs reflect the difference in lightcone sizes between DREaM (1 deg$^2$) and \textsc{UniverseMachine} and the SC SAM (2 deg$^2$).

\section{Discussion}
\label{sec:discussion}

The highly anticipated \emph{Roman Space Telescope} marks the beginning of a new era of \textit{deep-wide} surveys. \emph{Roman}'s wide field of view, sensitivity, and spatial resolution promise to deliver a new generation of \textit{deep-wide} surveys, which will reach depths comparable to the legacy ultra-deep \emph{Hubble} surveys (e.g. HUDF) or wide-field \emph{JWST} surveys while covering areas several times those of current generation wide-field galaxy surveys by space-based telescopes (e.g. CANDELS).
These future generation deep-wide surveys are expected to deliver more robust statistical constraints through detecting large populations of high-redshift galaxies. They are also expected to detect some very rare objects, including massive galaxies, luminous quasars, or perhaps even the first stars.
Semi-analytic models for galaxy formation have been shown to be a promising tool for interpreting legacy CANDELS observations \citep{Somerville2021} and forecasting for upcoming \emph{JWST} surveys \citep{Yung2022}.
The physically-motivated Santa Cruz SAM has been calibrated to observed constraints from the local universe and its performance in terms of reproducing the observed evolution of the high-redshift galaxy population has been rigorously tested in the series of \textit{Semi-analytic forecasts for JWST} paper series (e.g. \citealt{Yung2019}) and other works.
In this work, we present a set of multi-\sqdeg\ lightcones that have been created to prepare for future deep surveys with \emph{Roman} and other next generation wide-field survey telescopes, such as Euclid and Rubin.
These lightcones also include photometry for a large set of current generation of space- and ground-based instruments, as well as instruments used in legacy surveys. See Table \ref{table:catalogue_summary} for the full list of available photometry provided in the mock catalogues.

\subsection{The promise of deep-wide-field surveys}

The new generation of highly efficient wide-field survey instruments will redefine \textit{deep surveys}, as surveys of the size of a handful of telescope pointings will surpass the coverage of the widest `deep' surveys conducted with current generation instruments, and the largest `deep' surveys achievable with many pointings will reach at least tens of square degrees.
The many benefits of wide-field, high resolution space-based imaging with \emph{Hubble} that were discussed in \citet{Somerville2005} still hold today in a new era of high-redshift surveys with the next generation of wide-field survey instruments.
These wide-field surveys will be able to measure the stellar mass assembly history across cosmic time and disaggregate it by morphological type, large scale environment, and other galaxy properties.
The large galaxy sample expected from these surveys will also strengthen our understanding of the connection between the processes that regulate star formation on sub-galactic scales and the overall global trends in the star formation and mass assembly.
Furthermore, detecting galaxies with embedded accreting supermassive black holes forming in the early part of cosmic history will help reveal the relationship between star-forming galaxies and AGN, and the processes that regulate black hole feeding and growth.

CANDELS and and other large legacy surveys have addressed many of these questions and produced observational constraints that tremendously improved our understanding of galaxy formation in the context of cosmic evolution.
However, as observations push towards the very early Universe, some of these questions remain incompletely or imprecisely answered, limited by the capability of current generation instruments, which in turn limit feasible survey sizes and depths. Currently, only a few dozen objects at extreme redshifts (e.g. $z > 8$) have been detected to date. While \emph{JWST} is expected to find many more faint objects within the survey areas of familiar legacy surveys, it is not expected to find exotic objects such as massive, bright galaxies, as that would require a larger survey area.
\emph{Roman}, on the other hand, will deliver next generation wide-field surveys that are expected to efficiently detect large numbers of massive galaxies in the early Universe and will provide more robust statistics on the number counts of bright galaxies.

As demonstrated in Sections \ref{sec:variance} and \ref{sec:clustering}, the current generation surveys are susceptible to significant uncertainties arising from field to field variance, and these increase at higher redshift for a given galaxy mass or rest-frame luminosity.
\citet{Somerville2004} and \citet{Moster2011} have presented predictions for the cosmic variance expected in the legacy CANDELS fields at $0.5 < z < 4$.
As shown in \citet{Finkelstein2021}, based on a \twosqdeg\ lightcone presented in this work, the EGS field as observed by \emph{Hubble} and \emph{Spitzer} could be over-dense at $z\sim9$ relative to the cosmic mean.
Similarly, \citet{Yung2022} presented a controlled experiment with 40 realizations of of the order of $\sim1000$\,\sqdeg\ lightcones and showed that the field-to-field variance can be up to $\sigma_v \sim 0.40$ for survey reaching observed-frame IR magnitude in \emph{JWST}/NIRCam F200W $m_\text{F200W}\sim27$ at $z\sim9$.

Taking advantage of the new suite of \twosqdeg\ lightcones, we presented detailed predictions of field-to-field variance between $1 < z < 10$ as a function of survey area and depth.
We explored survey fields of different sizes ranging from approximately the size of UDF or a handful of \emph{JWST} NIRCam pointings to a single \emph{Roman} WFI pointing, and for depths that were reachable by legacy CANDELS surveys to those expected for ultra deep \emph{Roman} surveys.
This experiment presents quantitative predictions that enable informed estimates of the area needed to reduce field-to-field variance to a desired level for a population with specified intrinsic or observable properties.
For instance, Fig~\ref{fig:variance_byMag} shows that an ultra-deep \emph{Roman} survey (with area $\sim 1000$\,\sqarcmin\ reaching depth $m_{F184} <29$) will reduce $\sigma_v \sim 0.75$ at $z\sim9$ from a CANDELS-like survey (with $\sim200$\,\sqarcmin\ reaching depth $m_\text{IR} < 25$) to $\sigma_v \sim 0.10$. Obviously, our publicly available lightcones can be utilized to create similar estimates for any desired survey configuration that fits within their constraints on area and depth.

\subsection{Advantages of physics-based models}

Based on a well-established, versatile model for galaxy formation, we are able to provide a wide range of self-consistent, physically-backed predictions for high-resolution synthetic SEDs, multi-wavelength photometry across observatories, as well as the underlying physical properties and star formation history.
The semi-analytic modelling approach incorporates a wide range of physical processes and simultaneously and self-consistently explores their effects in a computationally efficient way. SAMs can also be used to explore the effects of varying the uncertain parameters that characterize these processes in current models, as shown in \citet{Yung2019, Yung2019a}. Achieving the dynamic range of the mock lightcones presented here (in galaxy mass/luminosity and area/volume) is well out of reach for current conventional numerical hydrodynamics simulations.
Our mock catalogues provide tens of millions of galaxies, which provide the statistically robust sample size required for forecasting future wide-field surveys.

On the other hand, although empirical models can very inexpensively fill large volumes with galaxies, they provide limited insights into the physical processes that shape galaxy properties. Moreover, they probably do not accurately represent the diversity of star formation histories in the galaxy population, which is important for producing a realistic suite of galaxy SEDs. These are important for designing and testing multi-instrument synergies and strategies. For example, narrow- and intermediate-band flux on HSC can efficiently refine the redshift estimates for Lyman-break galaxy candidates detected with broad-band dropout in VISTA/VIRCam \citep{Endsley2022}.
Combining flux measurements from filters on different instruments with a similar but slightly offset wavelength range can be used as a pseudo narrow band to refine redshift estimates (e.g. \emph{JWST}/NIRCam F150W and \emph{HST}/WFC3 F160W, see \citealt{Finkelstein2022a}), or two instruments with complementary wavelength coverage can be used (e.g. \emph{HST}/WFC3 Near-IR capability and \emph{Spitzer}/IRAC mid-IR capability, see \citealt{Finkelstein2021}). In addition, some of the most exciting applications of next generation wide-field surveys will be cross-correlation studies using galaxies detected via different multi-wavelength observational tracers. Current empirical models are not able to self-consistently predict, for example, the stellar mass, SFR, \ion{H}{I} and H$_2$ content, and dust content of galaxies. These different components of galaxies, as probed by different tracers such as UV, optical, and IR continuum emission, 21-cm emission, and sub-mm lines such as CO and [\ion{C}{II}], will ultimately provide very stringent constraints on and help to break degeneracies between different physical processes in galaxies. The same semi-analytic models presented here have also been used very successfully to predict some of these tracers \citep{Popping2019, Popping2019a, Yang2021a, Yang2022}.

\subsection{Our results in the context of other model predictions}
\label{sec:model_comp}

In this work, we performed a comparison between our predicted lightcones and two other mock lightcone suites constructed with the empirical models \textsc{UniverseMachine} \citep{Behroozi2019, Behroozi2020} and DREaM \citep{Drakos2021}.
We compared the stellar-to-halo mass ratio, bright and faint galaxy populations, and the angular correlation functions. Given that these lightcones are comparable in size and depth with the ones presented in this work, we are able to apply selection criteria for $m_\text{F184}$, $M_\text{UV}$, and $M_\text{*}$ similar to the ones adopted in earlier sections in this work.
While all three lightcones presented in this comparison share some modelling assumptions, including a $\Lambda$CDM cosmology (with comparable cosmological parameters), virial mass definition, and underlying IMF, it is worth noting that these models took very different approaches to modelling star formation histories, which is the main reason for the discrepancy seen in this comparison.
We note that although we do not make a direct comparison with them, related predictions for field-to-field variance have also been presented in the literature by \citet{Somerville2004, Moster2011, Trenti2008, Bhowmick2020, Endsley2020}.
We also note that there are also lightcones that have been generated based on hydrodynamic simulations \citep[e.g.][]{Snyder2017, Snyder2022, Kaviraj2017}, which are excellent for mock images and for studies that require realistic structural information for galaxies \citep[e.g.][]{Costantin2022, Garcia-Argumanez2022, Rose2022}. However, in general these lightcones span much smaller areas than the ones that we focus on in this work, and are therefore not included in the comparison.

Given the nearly identical assumptions that went into simulating the halo populations, the contribution to the difference among the models from the halo properties is expected to be minuscule.
Therefore, the stellar-to-halo mass ratios presented in Fig.~\ref{fig:models_SHMR} serve as direct diagnostics for the relationship between the stellar mass of galaxies and the mass of their host dark matter halos.
It is not surprising that the two empirical models, driven by a similar set of observational constraints, are in broad agreement.
Our model directly predicts the stellar-to-halo mass relation and the dispersion in it, and it is a result of the physical processes included in the model, including  merger-induced episodic star bursts and reduced star-formation activity in low-mass halos due to the ejection of gas by stellar feedback. While the model is calibrated to reproduce the observed stellar mass function \citep[e.g.][]{Bernardi2013} and SHMR \citep[e.g.][]{Rodriguez-Puebla2017} from abundance matching at $z\sim0$, it is not explicitly \textit{`tuned'} to match higher redshift constraints.
On the other hand, the empirical models are optimized to reproduce observed constraints over a wide range of redshifts, where the connection between stellar mass and halo mass is obtained by varying specific scaling relation(s) (e.g. SFR--$v_\text{max}$ and $f_\text{quenched}$--$v_\text{max}$ in \textsc{UniverseMachine}).
The SAM predicts very little redshift-evolution in the SHMR (for low-mass halos), while predictions from both empirical models increase monotonically towards high redshift. This is a direct consequence of the parametrization of the mass-loading of stellar driven winds with halo maximum circular velocity in the SAM.

Similarly, the comparison in Fig.~\ref{fig:count_models} helps demonstrate the differences between these models by breaking down the predicted galaxy populations, and sheds light on the differences seen in the predicted SHMR.
All three models produce similar predictions for the \emph{observable} quantity, the rest-UV luminosity function, as reflected in the top two panels of Fig.~\ref{fig:count_models}. However, the SAM accomplishes this with a different ratio of rest-UV light to stellar mass than the empirical models, which is due to the galaxies containing stellar populations with a different age and/or metallicity distribution, and/or a different assumed dust attenuation. The SAM self-consistently \emph{predicts} the joint age-metallicity distribution in each galaxy based on the star formation and assembly history computed from physical prescriptions, and the dust attenuation law is adjusted empirically to match the observations. The empirical models effectively construct star formation histories by matching to observational estimates of SFR in high redshift galaxies, and assume a relationship between stellar mass and metallicity that is obtained by extrapolating a lower redshift observational estimate.  The SAM generally predicts that there is more mass in stars at $4 \lesssim z \lesssim 7$ than the empirical models (bottom panel of Fig.~\ref{fig:count_models}).
As previously discussed in \citet{Yung2019a}, these predictions from our model are in tension with the evolution of $M_*/M_\text{h}$ over the interval $z\sim 4$--7 as derived from observations as presented in \citet{Finkelstein2015a}. See also fig.~4 in \citealt{Yung2019a} for the predicted stellar mass function from the Santa Cruz SAM and how it compares to the uncertainties in existing observations.
We note that the differences in the predicted number density of both massive and low-mass galaxies between the SAM and the two empirical models gradually decreases towards lower redshift at $z < 4$, which indicate the discrepancy in the predicted SHMR would narrow and eventually converge at $z \lesssim 1$.
It is important to note that there are very large uncertainties in the estimates of physical parameters such as stellar mass, SFR, and metallicity from observations, and these flow through to the empirical models. These constraints will be improved significantly by \emph{JWST} and other upcoming facilities including  \textit{Roman}.

The predicted angular correlation functions from the different lightcones are in remarkably good agreement when galaxies are selected via rest-UV luminosity. The large-scale separation ACF calculated for galaxies in the DREaM lightcone is slightly affected by the size of the lightcone, which is a quarter of the size of the \twosqdeg\ lightcones made with the Santa Cruz SAM and \textsc{UniverseMachine}.
Furthermore, differences across the ACF predicted by these different models are most noticeable in the small-scale separation, which is dominated by galaxies that reside within the same host halo (also referred to as the `one-halo term'). As already noted, the Santa Cruz SAM uses a rather different approach for modeling the positions of satellite galaxies within their host halos.

\subsection{Caveats, limitations, and uncertainties}

Given that the \twosqdeg\ lightcones presented in this work are constructed with the same physical model and tools as the wide-field and ultra-deep lightcones presented in \citet{Yung2022}, we refer the reader to that work (and the accompanying paper series) for discussion related to the construction of the lightcone and the predicted galaxies. Here, we focus on the caveats most closely related to the new results presented in this work.

Like many SAMs, the SC SAMs compute galaxy properties based on an input merger history. Thus, these models typically do not account for the effects of environment on scales larger than halos, such as the tidal interaction of halos in a high density environment (see these studies that account for impact from environment: \citet{Buck2019, Ayromlou2021, Kuschel2022}). Furthermore, the modelling of satellite galaxies also has some significant known issues and limitations. The SC SAM does not model the tidal or ram pressure stripping of satellites (which would impact their hot and cold gas, stellar bodies, and dark matter halos), and is known to produce satellite populations that are `over-quenched' relative to observations \citep{Kimm2009}.
A related issue is that the SC SAMs do not use the information on sub-halo evolution from the $N$-body simulations, but use a simple semi-analytic model to estimate the time required for satellite orbits to decay until they merge with the central galaxy, as well as the tidal stripping and destruction of satellites. Although the model has been tuned to produce good agreement with the satellite conditional mass functions and fractions (see Appendix \ref{appendix:c}), the predicted radial distribution of satellites does not match predictions from e.g. \textsc{UniverseMachine}. As a result, we have re-assigned the radial positions of the surviving satellites as described in Section~\ref{sec:models}.
In addition, in this work, in order to increase the dynamic range of our lightcones, we made use of halo merger histories constructed using the extended Press-Schechter formalism (see Section~\ref{sec:models}). These do not capture the known correlation between halo large scale environment (clustering) and mass accretion history, leading to small differences in the predicted clustering of galaxy populations selected by stellar mass or luminosity.

We also add that the sample selection throughout this work is idealized, but these catalogues provide a platform that can be used to include more realistic observational effects. While the effects of IGM extinction and ISM dust attenuation are included in the modelling pipeline, the photometry presented in this work does not include nebular emission from ionized H$_{\rm II}$ regions. It has been shown that strong emission line features in star forming galaxies can have an effect on the photometric redshift estimates from broad-band fluxes \citep[e.g.][]{Finkelstein2011, Finkelstein2022a, Larson2018}. In addition, we assumed that the observed magnitude includes the total intrinsic luminosity, while in real galaxy surveys, the extracted photometry for extended objects depends in a complex way on the galaxy size and light profile, the depth and S/N of the image, and the method used to extract the photometry.  In a companion paper, Bagley et al. (in preparation) add observational effects to these lightcones to more realistically explore the role of foreground contamination on the selection of high-redshift Lyman-break galaxies.
We add that gravitation lensing is another caveat that affects the recovered LF through magnification bias, and modulates the angular correlation function, but is outside of the scope of both this work and its companion work.
Also, in preparation for the \emph{JWST} Cosmic Evolution Early Release Survey (CEERS), we are creating noiseless mock images which then have \emph{JWST} instrument effects injected (Bagley and the CEERS Collaboration, in preparation).
Similar procedures can be used to generate realistic mock images for the \emph{Roman Space Telescope} and other facilities.

\section{Summary and conclusions}
\label{sec:snc}
We presented a suite of five \twosqdeg\ lightcones that spans $0 < z < 10$ and resolves galaxy populations down to  $\log(M_{*}/\text{M}_\odot) > 7$. We provide predictions for galaxies based on dark matter halos sourced from a cosmological $N$-body simulation and the physically motivated and well-established Santa Cruz semi-analytic model with configurations detailed in the \textit{Semi-analytic forecasts for JWST} project. Taking advantage of the large area and large number of sample galaxies, we present forecasts that highlight the potential of future surveys that will be orders of magnitude wider than current generation observations of comparable depth. In this work, we focus on predictions that utilise the spatial distribution and redshift evolution for galaxies available in these lightcones.

We summarize our main conclusions below.
\begin{enumerate}
    \item We present the predicted evolution of object counts and number density as a function of redshift, as well as the contribution of these galaxies to the cosmic SFR.

    \item We predict that the uncertainty due to field to field variance for deep surveys ($m_\text{F184,lim}\sim29$) can be reduced to about $\sim10$\% at $z\sim 6$--10 for a 1000\,\sqarcmin\ survey. For brighter limits ($m_\text{F184,lim}\sim24$), the uncertainty due to field-to-field variance for such a survey is $25$--$50$ per cent at this survey area.

    \item We predict that for magnitude limits typical of moderate depth \emph{Roman} surveys ($m_\text{F184,lim}\sim28$), the uncertainty on two-point correlation function estimates in 0.5 deg$^2$ fields at $z\sim 4$--4.5 for separation $\sim10$ to 100 arcsecond due to field-to-field variance is $\sim15$ per cent. Surveys with an area of a few \sqdeg\ should be sufficiently large for field-to-field variance to be a sub-dominant source of uncertainty.

    \item We compared our predictions with two other lightcone simulations from the literature that populated halos with galaxies using empirical models. Our physics-based models predict a significantly higher stellar mass to halo mass relation at $z\sim 4$ compared with the empirical models, although all three models match the observed rest-UV luminosity function. All three models make similar predictions for the angular correlation functions at $z\sim 4$--9.
\end{enumerate}

The results presented in this paper are intended as just a few examples of the predictions that can be extracted from these lightcones. All of the lightcones are made publicly available so that the community can exploit them for numerous additional applications.

\section*{Acknowledgements}

The authors of this paper would like to thank David Spergel, Stephen Wilkins, Mark Dickinson, Austen Gabrielpillai, Shengqi Yang, Nicole Drakos, Stephen Wilkins, and Madeline Marshall for useful discussions.
We also thank the members of the \textit{Roman Space Telescope} Cosmic Dawn Science Investigation Team and the Cosmological Advanced Survey Telescope for Optical and uv Research (CASTOR) Science Team for utilizing the pre-production results and providing feedback that improved this work.
We thank the anonymous referee for the constructive comments that improved this work.
The simulations and data products for this work were run on NASA computing machines, \textit{astera} and \textit{seliana}, managed by the Office of Scientific Computing at NASA Goddard Space Flight Center.
We thank the Center for Computational Astrophysics (CCA) and the Scientific Computing Core (SCC) at the Flatiron Institute for hosting the data associated with this work on the data release portal \textit{Flathub}.
We warmly thank Dylan Simon and Elizabeth Lovero for coordinating the data release portal and project website, and Rebecca Sesny for the creation of the project website.
AY is supported by an appointment to the NASA Postdoctoral Program (NPP) at NASA Goddard Space Flight Center, administered by the Oak Ridge Associated Universities under contract with NASA.
AY also thanks the CCA for hospitality during the creation of this work.
RSS acknowledges support from the Simons Foundation.

Use of the \twosqdeg\ mock catalogues presented in this work should reference both this work and \citealt{Somerville2021} for the construction of the simulated lightcones, and reference \citet{Somerville2015} for the Santa Cruz semi-analytic model and \citet{Yung2019} for model calibration and validation against existing observational constraints. In addition, reference \citet{Yung2022} if \emph{JWST} NIRCam broad- and medium-band photometry is used.

%%%%%%%%%%%%%%%%%%%%%%%%%%%%%%%%%%%%%%%%%%%%%%%%%%
\section*{Data Availability}

The data underlying this paper are available in the Data Product Portal hosted by the Flatiron Institute at \url{https://www.simonsfoundation.org/semi-analytic-forecasts/} and the Flatiron Institute Data Exploration and Comparison Hub (Flahub, \url{https://flathub.flatironinstitute.org/group/sam-forecasts}).



\bibliographystyle{mnras}
\bibliography{library.bib}
%\bibliography{/Users/layung/Publications/Bibliography/SAM_Roman.bib}

%\begin{thebibliography}{99}
%\end{thebibliography}

%%%%%%%%%%%%%%%%%%%%%%%%%%%%%%%%%%%%%%%%%%%%%%%%%%

%%%%%%%%%%%%%%%%% APPENDICES %%%%%%%%%%%%%%%%%%%%%
%\newpage
\appendix

\section{Summary of available columns in the lightcones}
\label{appendix:a}
\setcounter{table}{0} \renewcommand{\thetable}{A\arabic{table}}
In Table \ref{table:catalogue_summary}, we summarize the predicted quantities available in the mock catalogues. In addition to basic coordinate information, such as RA, Dec, and redshift, we provide a comprehensive set of halo properties and galaxy physical properties. We also provide rest- and observed-frame photometry for a wide variety of space- and ground-based telescopes, including Galaxy Evolution Explorer (GALEX), Sloan Digital Sky Survey (SDSS), Advanced Camera for Surveys (ACS) and Wide Field Camera 3 (WFC3) on-board \emph{Hubble}, Infrared Array Camera (IRAC) on-board \emph{Spitzer}, United Kingdom Infrared Telescope (UKIRT), Visible and Infrared Survey Telescope for Astronomy (VISTA),  Dark Energy Camera (DECam) of the Dark Energy Survey, \emph{JWST} NIRCam broad- and medium-bands, \emph{Roman} WFI, Euclid Observatory, and Rubin Observatory (formerly known as LSST).

While we present photometry for a large collection of space- and ground-based telescopes in this unified catalogue to maximize convenience, we add that the development and verification of some of these results are from past studies. Here we give a general summary for the references of past studies. Reference \citet[][for $z = 0$ to 6]{Somerville2021} and \citet[][for $z = 4$ to 10]{Yung2019, Yung2019a} for predicted physical properties of galaxies.
Reference this work for \emph{Roman}, \emph{Euclid}, \emph{Rubin} (labelled as LSST), and DECam.
Reference \citet{Yung2019, Yung2022} for \emph{JWST} photometry.
Reference \citet{Somerville2021} for standard rest-frame filters, \emph{HST}, \emph{Spitzer}, SDSS, GALEX, UKIRT, VISTA, and other filters used in the CANDELS survey.

\begin{table*}
    \centering
    \caption{List of all predicted quantities available in the mock galaxy catalogues. We provide the columns (in the ASCII format catalogue `\texttt{lightcone.dat}'), brief descriptions, and units associated with these quantities. For rest- and observed-frame, we list the intrinsic, unattenuated photometry filter sets and filter bands available (e.g. \texttt{UV1500\_rest}, \texttt{UV2300\_rest}, and \texttt{UV2800\_rest}), and for each band we also provide unattenuated luminosity of the bulge (\texttt{*\_bulge}, e.g. \texttt{UV1500\_rest\_bulge}) and with dust attenuation (\texttt{*\_dust}, e.g. \texttt{UV1500\_rest\_dust}). The wide-field and ultradeep lightcones presented in \citet{Yung2022} share the exact same format and columns.}
    \label{table:catalogue_summary}
    \begin{tabular}{lllll}
        \hline
        Category                    & Column  & Quantity                        & Descriptions                                               & Units                  \\
                                    \hline
                                    & 0       & \texttt{halo\_id\_nbody}        & \multicolumn{2}{l}{id copied from n-body simulations, unique with lightcone}        \\
                                    & 1       & \texttt{gal\_id}                & \multicolumn{2}{l}{id assigned to galaxies, unique within each halo}                \\
                                    & 2       & \texttt{gal\_type}              & 0 = central, 1 = satellite                                 &                        \\
                                    & 3       & \texttt{z\_nopec}               & cosmological redshift without peculiar velocity            &                        \\
                                    & 4       & \texttt{redshift}               & cosmological redshift with peculiar velocity               &                        \\
                                    & 5       & \texttt{ra}                     & right ascension                                            &  Deg                   \\
                                    & 6       & \texttt{dec}                    & declination                                                &  Deg                   \\
                                    \hline
        Halo Properties             & 7       & \texttt{m\_vir}                 & virial mass                                                &  $10^{10}$ M$_\odot$   \\
                                    & 8       & \texttt{V\_vir}                 & virial velocity                                            &  km s$^{-1}$           \\
                                    & 9       & \texttt{r\_vir}                 & virial radius                                              &  Mpc                   \\
                                    & 10      & \texttt{c\_NFW}                 & concentration parameter in NFW profile                     &                        \\
                                    & 11      & \texttt{spin}                   & spin parameter of halo                                     &                        \\
                                    & 12      & \texttt{mstar\_diffuse}         & mass of diffuse stellar population in halo                 &  $10^{10}$ M$_\odot$   \\
                                    & 13      & \texttt{m\_hot\_halo}           & hot gas mass in halo                                       &  $10^{10}$ M$_\odot$   \\
                                    & 14      & \texttt{Z\_hot\_halo}           & hot gas metallicity in halo                                &  Z$_\odot$             \\
                                    \hline
        Galaxy Properties           & 15      & \texttt{v\_disk}                & velocity of disk                                           &  km s$^{-1}$           \\
                                    & 16      & \texttt{r\_disk}                & radius of disk                                             &  kpc                   \\
                                    & 17      & \texttt{sigma\_bulge}           & velocity dispersion of bulge                               &  km s$^{-1}$           \\
                                    & 18      & \texttt{rbulge}                 & bulge radius                                               &  kpc                   \\
                                    & 19      & \texttt{mhalo}                  & halo mass                                                  &  $10^{10}$ M$_\odot$   \\
                                    & 20      & \texttt{mstar}                  & stellar mass                                               &  $10^{10}$ M$_\odot$   \\
                                    & 21      & \texttt{mcold}                  & cold gas mass                                              &  $10^{10}$ M$_\odot$   \\
                                    & 22      & \texttt{mbulge}                 & bulge mass                                                 &  $10^{10}$ M$_\odot$   \\
                                    & 23      & \texttt{mbh}                    & black hole mass                                            &  $10^{10}$ M$_\odot$   \\
                                    & 24      & \texttt{maccdot}                & black hole accretion rate                                  &  M$_\odot$ yr$^{-1}$   \\
                                    & 25      & \texttt{maccdot\_radio}         & black hole accretion rate (radio mode)                     &  M$_\odot$ yr$^{-1}$   \\
                                    & 26      & \texttt{Zstar}                  & stellar metallicity                                        &  Z$_\odot$             \\
                                    & 27      & \texttt{Zcold}                  & cold gas metallicity                                       &  Z$_\odot$             \\
                                    & 28      & \texttt{mstardot}               & instantaneous SFR                                          &  M$_\odot$ yr$^{-1}$   \\
                                    & 29      & \texttt{sfr\_ave}               & SFR averaged over 100 Myr                                  &  M$_\odot$ yr$^{-1}$   \\
                                    & 30      & \texttt{meanage}                & mass-weighted mean stellar age                             &  Gyr                   \\
                                    & 31      & \texttt{tmerge}                 & time elapsed since last merger                             &  Gyr                   \\
                                    & 32      & \texttt{tmajmerge}              & time elapsed since last major merger                       &  Gyr                   \\
                                    & 33      & \texttt{cosi}                   & angle of inclination                                       &                        \\
                                    & 34      & \texttt{tauV0}                  & extinction optical depth in the $V$-band                   &                        \\
                                    & 35      & \texttt{maccdot\_BH}            & black hole accretion rate                                  &  M$_\odot$ yr$^{-1}$   \\
                                    & 36      & \texttt{sfr10myr}               & SFR averaged over 10 Myr                                   &  M$_\odot$ yr$^{-1}$   \\
                                    & 37      & \texttt{mstarold}               & mass of stars older than 1 Gyr                             &  Z$_\odot$             \\
                                    & 38      & \texttt{ageold}                 & mass-weighted mean age of stars older than 1 Gyr           &  Gyr                   \\
                                    & 39      & \texttt{zstarold}               & metallicity of stars older than 1 Gyr                      &  Z$_\odot$             \\
                                    \hline
                                    &         & filter set                      & \multicolumn{2}{l}{available bands (all magnitudes are given in the AB system)}     \\
                                    \hline
        Rest-frame photometry       & 40-48   & \texttt{\{\,\,\,\,\,\}\_rest}   & \texttt{UV1500, UV2300, UV2800}                                                    &\\
                                    & 49-72   & \texttt{\{\,\,\,\,\,\}\_rest}   & \texttt{U, B, V, R, I, J, H, K}                                                    &\\
%                                    \hline
        Observed-frame photometry   & 73-78   & \texttt{galex\_\{\,\,\,\,\,\}}  & \texttt{FUV, NUV}                                                                  &\\
                                    & 79-93   & \texttt{sdss\_\{\,\,\,\,\,\}}   & \texttt{u, g, i, r, z}                                                             &\\
                                    & 94-108  & \texttt{acs\{\,\,\,\,\,\}}      & \texttt{f435w, f606w, f775w, f814w, f850lp}                                        &\\
                                    & 109-123 & \texttt{wfc3\{\,\,\,\,\,\}}     & \texttt{f275w, f336w, f105w, f125w, f160w}                                         &\\
                                    & 133-141 & \texttt{UKIRT\_\{\,\,\,\,\,\}}  & \texttt{J, H, K}                                                                   &\\
                                    & 142-147 & \texttt{irac\_\{\,\,\,\,\,\}}   & \texttt{ch1, ch2}                                                                  &\\
                                    & 148-171 & \texttt{NIRCam\_\{\,\,\,\,\,\}} & \multicolumn{2}{l}{\texttt{F070W, F090W, F115W, F150W, F200W, F277W, F356W, F444W}} \\
                                    & 172-204 & \texttt{NIRCam\_\{\,\,\,\,\,\}} & \multicolumn{2}{l}{\texttt{F140M, F162M, F182M, F210M, F250M, F335M, F360M, F410M}} \\
                                    &         &                                 & \multicolumn{2}{l}{\texttt{F430M, F460M, F480M}}                                    \\
                                    & 205-216 & \texttt{Euclid\_\{\,\,\,\,\,\}} & \multicolumn{2}{l}{\texttt{VIS, Y, J, H}}                                           \\
                                    & 217-240 & \texttt{Roman\_\{\,\,\,\,\,\}}  & \multicolumn{2}{l}{\texttt{F062, F087, F106, F129, F146, F158, F184, F213}}         \\
                                    & 241-258 & \texttt{LSST\_\{\,\,\,\,\,\}}   & \multicolumn{2}{l}{\texttt{u, g, r, i, z, y}}                                       \\
                                    & 259-276 & \texttt{DECam\_\{\,\,\,\,\,\}}  & \multicolumn{2}{l}{\texttt{u, g, r, i, z, Y}}                                       \\
                                    & 277-279 & \texttt{NEWFIRM\_K\_atm}        &                                                                                    &\\
                                    & 280-294 & \texttt{VISTA\_\{\,\,\,\,\,\}}  & \multicolumn{2}{l}{\texttt{z, Y, J, H, Ks}}                                         \\
        \hline
    \end{tabular}
\end{table*}

\section{Code snippet for calculating volume in a lightcone slice}
\label{appendix:b}
\setcounter{table}{0} \renewcommand{\thetable}{B\arabic{table}}

This simple piece of code is very handy for calculating the comoving volume for a slice of the lightcone. This is essential to calculate the volume-averaged quantities presented in this work.
We use the \texttt{comoving\_volume} function from \texttt{astropy} to calculate the comoving volume of the whole universe at two given redshift \texttt{z1} and \texttt{z2}, and obtain the comoving volume of the universe between \texttt{z1} and \texttt{z2}. We then multiple the full-sky co-moving volume by the fraction of the surveyed area (in square arcminutes) over $4\pi$ steradian $\approx$ 148510656 \sqarcmin\ to obtain the volume within the surveyed area.

\begin{verbatim}
from astropy.cosmology import FlatLambdaCDM
cosmo = FlatLambdaCDM(Om0=0.307, Ob0=0.048, H0=67.8)

def volume_estimate(z1, z2, sqarcmin):
    volume_z1 = cosmo.comoving_volume(z1)
    volume_z2 = cosmo.comoving_volume(z2)
    sky_fraction = sqarcmin/148510656.

    vol = (volume_z2-volume_z1).value*sky_fraction
    return abs(vol) ### return volume in cMpc^3
\end{verbatim}

\section{Satellite galaxy comparison}
\label{appendix:c}
\begin{figure*}
    \includegraphics[width=2\columnwidth]{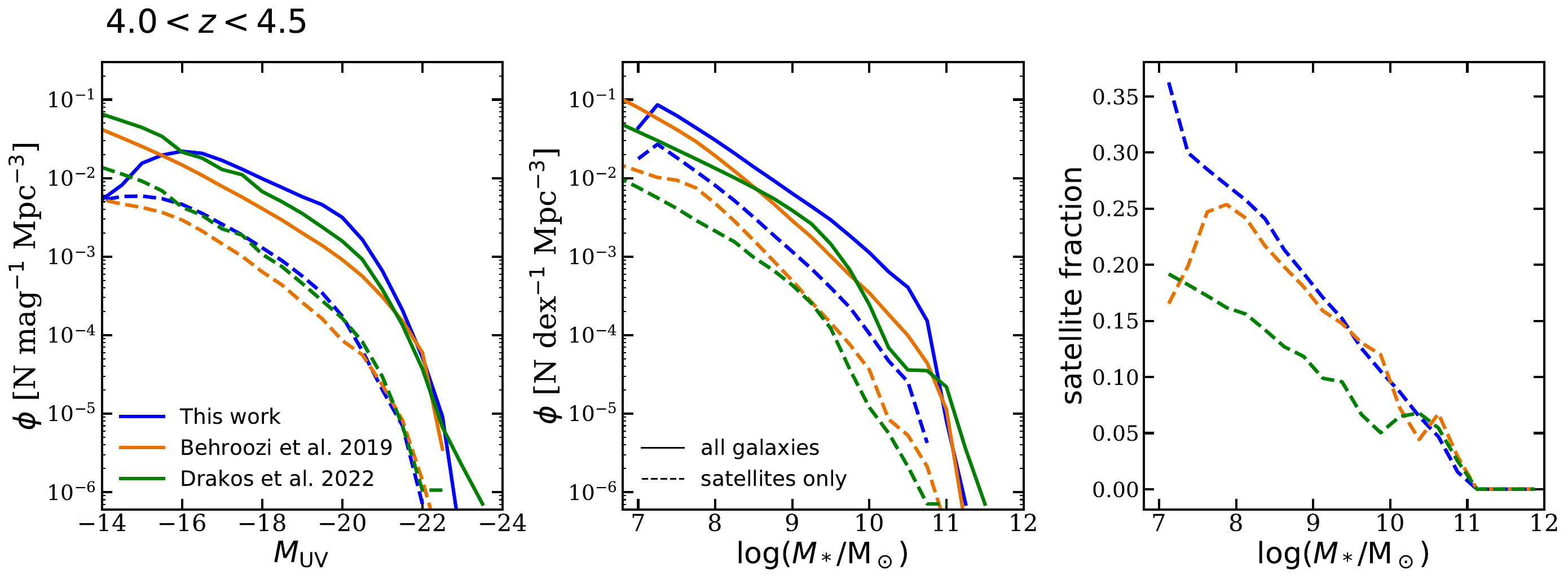}
    \caption{
        UV luminosity functions (\textit{left}) and stellar mass functions (\textit{middle}) predicted from one of the \twosqdeg\ lightcones, the DREaM mock catalogue, and a \textsc{UniverseMachine} \twosqdeg\ mock catalogue. We show distribution functions for all galaxies and satellite galaxies with solid and dashed lines, respectively. We also show the satellite fraction as a function of stellar mass (\textit{right}).
    }
    \label{fig:satellite_distfn_frac}
\end{figure*}

\setcounter{table}{0} \renewcommand{\thetable}{C\arabic{table}}
In this Appendix, we provide a set of comparisons across the Santa Cruz SAM, \textsc{UniverseMachine}, and DREaM for galaxies between $4.0 < z < 4.5$. Satellites are identified with \texttt{gal\_id!=1} for the Santa Cruz SAM, \texttt{UPID!=-1} for \textsc{UniverseMachine}, and \texttt{hostID!=-1} for DREaM.
In Fig.~\ref{fig:satellite_distfn_frac}, we show predicted UV luminosity functions and stellar mass functions for all galaxies and satellite galaxies presented in these lightcones. In the last panel, we also show the fraction of satellite galaxies as a function of stellar mass found in these models. The overall UV luminosity functions and stellar mass functions match fairly well between the three models (within the observational errors). The fraction of galaxies that are satellites as a function of stellar mass matches quite well between the Santa Cruz SAM and \textsc{UniverseMachine}, while DREaM produces satellite fractions that are slightly lower at low masses.

\begin{figure*}
    \includegraphics[width=2\columnwidth]{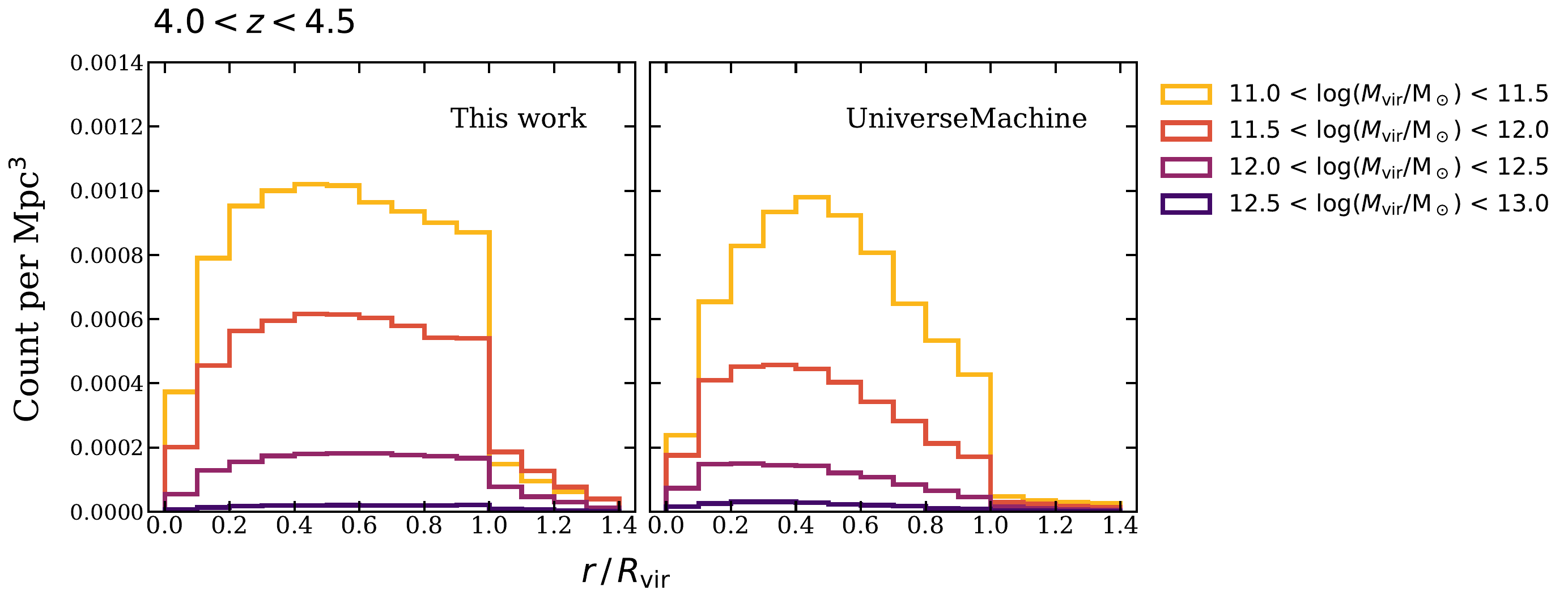}
    \caption{
        Histograms for the volume-averaged counts for predicted satellite galaxies binned by their radial distances normalized to the virial radii of their host halos, $r/R_\text{vir}$, in one of the 2-\sqdeg\ lightcones presented in this work (\textit{left}) and in the \textsc{UniverseMachine} GOODS-S lightcone (\textit{right}) in virial mass bins with $\Delta \log(M_\text{vir}/\text{M}_\odot) = 0.5$ for satellite galaxies encompassed in halos with mass $\log(M_\text{vir}/\text{M}_\odot) = 11.0$ to 13.0.
    }
    \label{fig:satellite_distances}
\end{figure*}

In Fig.~\ref{fig:satellite_distances}, we show histograms for the number counts of satellite galaxies, normalized to the volume of the lightcones, binned by their radial distances normalized to the virial radii of their host halos, $r/R_\text{vir}$, for the Santa Cruz SAM and \textsc{UniverseMachine}. These show reasonably good agreement.

%%%%%%%%%%%%%%%%%%%%%%%%%%%%%%%%%%%%%%%%%%%%%%%%%%

% Don't change these lines
\bsp% typesetting comment
\label{lastpage}
\end{document}